\documentclass[fleqn,usenatbib]{mnras}

\usepackage{newtxtext,newtxmath}
\usepackage{adjustbox, tabularx, makecell}
\usepackage{multirow}
\usepackage[greek, english]{babel}
\usepackage{alphabeta}
\usepackage{textgreek}

\usepackage[T1]{fontenc}

\DeclareRobustCommand{\VAN}[3]{#2}
\let\VANthebibliography\thebibliography
\def\thebibliography{\DeclareRobustCommand{\VAN}[3]{##3}\VANthebibliography}

\usepackage{graphicx}	
\usepackage{amsmath}	

\title[Observational constraints of diffusive dark-fluid cosmology]{Observational constraints of diffusive dark-fluid cosmology}

\author[S. Sahlu et al]{
Shambel Sahlu,$^{1,2}$\footnote{\thanks{sahlushambel@gmail.com}}
Upala Mukhopadhyay,$^{3}$ 
Remudin R. Mekuria$^{4}$ 
and Amare Abebe$^{1,2}$ 
\\
$^{1}$Centre for Space Research, North-West University, Potchefstroom, South Africa\\
$^{2}$National Institute for Theoretical and Computational Sciences (NITheCS), South Africa\\
$^{3}$Department of Physics and Materials Science, University of Luxembourg, L-1511 Luxembourg City, Luxembourg\\
$^{4}$Faculty of Engineering and Computer Science, Ala-Too International University, Bishkek, Kyrgyzstan\\
}

\date{Accepted XXX. Received YYY; in original form ZZZ}

\pubyear{\the\year{}}

\begin{document}
\label{firstpage}
\pagerange{\pageref{firstpage}--\pageref{lastpage}}
\maketitle

\begin{abstract}

{In this manuscript, we investigate late-time cosmology and the evolution of cosmic structures using an interacting dark fluid model in which dark matter (DM) and dark energy (DE) interact through a diffusive mechanism. To provide a comprehensive understanding, we derive the background evolution and perturbation equations within this model and obtain cosmological parameters through MCMC simulations. We use recent measurements for statistical analysis and constrain the parameters \(H_0\) in km/s/Mpc, \(\Omega_m\), \(r_d\), \(M\), \(\sigma_8\), \(S_8\), and the interaction term $Q_{dm}$. From the constrained values of $Q_{dm}$, we show that the diffusive model is a promising alternative DE model, capable of driving late-time cosmic acceleration due to energy exchange from DM to DE. State-finder diagnostics indicate that the model behaves like a Chaplygin gas when energy transfers from DM to DE during the Universe’s expansion. We also investigate the growth of density contrast, finding \(\delta_m(z)\gg\delta_{de}(z)\), which highlights the dominant role of DM in structure formation. Redshift space distortion and growth rate analysis show that minor deviations from \(\Lambda\)CDM at low redshifts, with larger differences at higher redshifts, indicate the impact of energy diffusion on early structure growth. Finally, we perform a detailed statistical analysis, including \({\mathcal{L}(\hat{\Theta}|data)}\), \(\chi^2\), \(\rm{AIC}\), and \(\rm{BIC}\), which strongly supports the proposed diffusive dark-fluid model.
}
\end{abstract}


\section{Introduction}\label{sec:intro}
 The mysterious matter and energy in the Universe, referred to as DM and DE, respectively, account for a whopping 95\% of the universe's total content. The nature of these dark components of the Universe is not properly understood, but several candidates in the literature, including unified dark-fluid models, have been proposed to describe them and their effect on astrophysics and cosmology.  On the DM side, the most commonly studied candidates include Weakly Interacting Massive Particles (WIMPS) \citep{bergstrom2009dark, colafrancesco2006multi, Mekuria:2017TI, mekuria2017multi, Mekuria_2019} or some astrophysical modification of gravity such as Modified Newtonian Dynamics (MOND)\citep{milgrom1983modification} among many others, whereas on the DE side, the cosmological constant $\Lambda$ \citep{carroll2001cosmological} is perhaps the simplest addition to the standard cosmological model needed to explain most of the observed data. However, there are some serious issues associated with the cosmological constant, such as the eponymous {\it cosmological constant problem} \citep{weinberg1989cosmological}, the coincidence problem \citep{velten2014aspects}, the Hubble tension \citep{DiValentino:2020zio}, the $S_8$ tension \citep{DiValentino:2020vvd}, the JWST high redshift massive galaxy tensions \citep{Labbe:2022ahb}, etc., which makes the choice less attractive.  That is why there are currently a plethora of other alternatives to explain current cosmological observations, such as modifications to the gravitational theory itself (see e.g.\cite{saridakis2021modified,bamba2012dark} and references therein), an evolving $\Lambda$ \citep{peebles1999evolution,peebles1988cosmology}, deviations from the standard homogeneous (see \cite{bolejko2011inhomogeneous,krasinski1997inhomogeneous} and references therein) and isotropic Universe (such as the various Bianchi cosmological models) assumption, or some form of a combination of these, among others.\\
 \\
 Another aspect that has gained much traction recently is the interaction of DM and DE \citep{bolotin2015cosmological,zimdahl2005interacting,Mukhopadhyay:2019jla,Mukhopadhyay:2020bml,chakraborty2025desi}. Such an approach is interesting because it has the potential to explain the cosmological and coincidence problems, the Hubble tension, and the $S_8$ discrepancy \citep{nunes2021arbitrating}. Some theoretical conditions that such interactions have to satisfy for physical viability have been studied in \cite{vanide}.
Diffusive dark-fluid cosmology proposes that DM and DE are different phases of a unified dark fluid, with a diffusion process that allows energy or mass to flow between them \cite{arbey2006dark}.  In recent decades, interacting dark-fluid models have garnered significant attention in cosmology, with extensive studies exploring their various aspects. In particular, investigations of dark-fluid interactions have aimed to address the cosmic coincidence problem \cite{wang2016dark}. Furthermore, cosmological evolution is broadly discussed in \cite{bolotin2015cosmological, sharov2017new} in the interaction between DE and DM.  Detailed thermodynamic stability analysis of diffusive dark fluid has been studied in recent work \cite{maity2019Universe}. The work in \cite{wang2007interacting,zhai2023consistent} reported the effect of the interaction between DM and DE on the formation of large-scale structures.  This effect is clearly shown at the lowest pole $l$ of the Cosmic Microwave Background (CMB) spectrum.
\\
\\The diffusive dark fluid is also proposed to alleviate the cosmological tensions as presented in \cite{nunes2016new,yang2014cosmological,calogero2013cosmology}. Earlier in the Universe, this fluid behaves like DM, helping to form structures, while later it acts like DE, driving cosmic acceleration \cite{zhang2007}. The model modifies standard cosmological equations and offers a unified approach to explaining DM and energy. While it simplifies the overall picture, it requires careful tuning of parameters to match observations like the CMB and galaxy surveys. Understanding the rate and underlying causes of this acceleration is crucial for forecasting the ultimate fate of the Universe, whether it will expand indefinitely or face a ``big rip". Furthermore, the discrepancy in Hubble constant measurements suggests the need for new physics or highlights potential errors in current observational methods. This accelerated expansion also challenges Einstein's theory of general relativity (GR), encouraging cosmologists to propose beyond GR theory for more accurate explanations without the need for a cosmological constant or DE. For instance, { the CG model \citep{bento2002generalized, fabris2002mass, sahlu2019chaplygin, sahlu2023confronting,Bento:2002ps,Bento:2002yx,Bento:2004uh} suggests that an exotic matter with negative pressure could serve as both DM and DE, while the phantom model \citep{elizalde2004late, csaki2006accelerated} proposes a more negative pressure form of DE, leading to accelerating acceleration and super-exponential expansion. Additionally, various modified gravity theories, such as  $f(R)$ gravity, based on the Ricci scalar $R$ \cite{bahamonde2017deceleration,abebe2012covariant,abebe2013large}, $f(Q)$ gravity, where $Q$ is a non-metric \cite{solanki2022accelerating, sahlu2025constraining,sahlu2025structure}, $f(T)$ gravity, involving the torsion scalar $T$ \cite{myrzakulov2011accelerating, cardone2012accelerating, sahlu2019accelerating,sahlu2020scalar} just mentioned few,  have been proposed to explain the accelerating expansion of the Universe.}
\\
\\
In the current work, we investigate the background evolution and cosmological perturbations within the framework of diffusive dark fluid interaction. Our analysis utilizes various recent cosmological datasets, including: (a) BAO distance and correlation measurements from the DR2 data release of the Dark Energy Spectroscopic Instrument Survey (\textit{DESI DR2 BAO}) \cite{andrade2025validation, abdul2025desi}; (b) cosmic chronometers (\textit{CC}) data \cite{moresco2020setting, qi2023model}, which measure the Hubble parameter $H(z)$ based on the relative ages of massive, early-time, passively evolving galaxies; (c) Type Ia supernovae (SNIa) datasets, specifically: 
i) the PantheonPlus+ sample \cite{brout2022pantheon+}, consisting of 1701 light curves of 1550 distinct SNIa across a redshift range of $z \in [0.001, 2.26]$ (\textit{PPS}); 
ii) the \textit{DESY5} dataset \cite{collaboration2024dark}, a photometrically classified SNIa with redshifts between $0.1$ and $1.13$, supplemented by 194 low-redshift SNe Ia spanning $0.025 < z < 0.1$; and 
iii) the latest Union3 compilation, which includes 2,087 cosmologically useful SNIa from 24 datasets \cite{rubin2023union}. We also consider, the redshift-space distortion data (\text{RSD}) and the growth rate \textit{f},  from the VIMOS Public Extragalactic Redshift Survey (VIPERS) and SDSS collaborations.  To achieve this, we focus on the theoretical framework of late-time and perturbations, as investigated by considering the diffusive dark fluid model that unifies DM and DE through a diffusive interaction, leading to a coupled evolution that drives cosmic evolution. For better analysis, we use the combined datasets: \textit{DESI DR2 BAO + CC + DESY5+RSD+f},  \textit{DESI DR2 BAO + CC +Union3 +RSD+f} and  \textit{ PPS  + CC + DESY5+RSD+f}  for constraining the cosmological parameters:  \(H_0\), \(\Omega_m\), \(r_d\), \(M\), \(\sigma_8\), \(S_8\) and the interaction term $Q_{dm}$ through the MCMC method. Then we highlighted the values of the Hubble parameter $H_0$ and matter clustering $S_8$ to emphasize the possibilities of the diffusive model in mitigating cosmological tensions. \textit{However, addressing the cosmological tensions is beyond the current manuscript.} After constraining the parameters through the MCMC method, different background cosmological quantities namely:  the acceleration parameter $q(z)$, the effective equation of state parameter $w_{eff}(z)$, the Hubble parameter $H(z)$, and the distance modulus $\mu(z)$ have been illustrated in the diffusive dark-fluid model. The state finder diagnostic plots, \( q \) versus \( r \) and \( s \) versus \( r \)  are also highlighted for the case of positive and negative \( Q_{dm} \) (energy flow from DM to DE and vis versa), offering insights into cosmic expansion and the diffusive dark-fluid model compared to the standard $\Lambda$CDM model. The positive \( Q_{dm} \) case corresponds to a quintessence-like phase, resulting in a slower expansion, while the negative \( Q_{dm} \) case resembles the CG model, leading to faster expansion in the late-time Universe than $\Lambda$CDM model. We are also devoted to investigating the $Om(z)$ diagnostic, which plays a significant role in cosmology by offering a model-independent approach to differentiate between various cosmological models, particularly about the nature of DE and the expansion history of the Universe. Its primary function is to help distinguish the standard $\Lambda$CDM model from the alternative DE, such as the diffusive model. In the case of $\Lambda$CDM  where DE is modeled as a cosmological constant, $Om(z)$ is expected to remain nearly constant across different redshifts. {On the whole, the diffusive dark-fluid model indicates a Universe in which DE may not be a straightforward cosmological constant but rather a dynamically changing phenomenon across time. This might lead to an improved understanding of cosmic acceleration and challenge the simplicity of the traditional $\Lambda$CDM model.}\\ \\
We also employ cosmic perturbation theory as a tool to understand how the large-scale structures we observe today have evolved and expanded as a result of gravitational instabilities in the early Universe. In the standard metric perturbation theory pioneered by Lifshitz \cite{lifshitz46} and later refined by Bardeen \cite{ bardeen80} and Kodama and Sasaki \cite{KS1984}, one usually starts by perturbing away from a homogeneous and isotropic background metric. The 1+3-covariant and gauge-invariant perturbation formalism \cite{hawking66,olson76,eb89,dbe92a,bed92,dbe92b}, on the other hand, starts by defining covariant and gauge-invariant gradient variables that define fluctuations in a given cosmological quantity (such as the energy density and the volume expansion) \cite{eb89,carloni2008, aadd} without specifying the background metric from the start.\footnote{See~\citep{gidelew2009} and the references therein for more details on this and the pros and cons of the two approaches of cosmological perturbation theory.} This paper also focuses on analysing linear cosmological perturbations to explore how the diffusive dark fluid model impacts the development of large-scale structures in the Universe. We have implemented the 1+3 covariant formalism introduced by \cite{ hawking66, olson76,eb89,dbe92a,bed92,dbe92b,ellis1989covariant,ellis1990density}, which distinguishes between time and space through an observer's four-velocity \(u^a\) and the projection tensor \(h_{ab}\). This approach streamlines the examination of spacetime and matter dynamics. Important quantities in this framework include the expansion scalar \( \Theta \), shear \( \sigma_{ab} \), and vorticity \( \omega_{ab} \), which characterize cosmic expansion, shape distortions, and rotational effects. This formalism also allows for the decomposition of the energy-momentum tensor, facilitating the analysis of perturbations, structure formation, the cosmic microwave background, and the expansion of the Universe, particularly within modified gravity models.
The evolution equation for the density contrast \(\delta_m(z)\) is examined using the 1+3 formalism across various gravity theories including general relativity \cite{dunsby1992cosmological, dunsby1992cosmological}, $f(R)$ gravity \cite{abebe2013large, abebe2012covariant}, $f(T)$ gravity \cite{sahlu2020scalar, sami2021covariant}, $f(Q)$ gravity \cite{sahlu2025structure, sahlu2025constraining}, the CG model \cite{sahlu2023confronting}, the $f(R, L_m)$ gravity \cite{sahlu2024cosmology}  and scalar-tensor theories \cite{ntahompagaze2018study, ntahompagaze2020multifluid} to study the growth of cosmic structures. In the current work,  we also implement the same manner for the diffusive model to study the scalar perturbations, which leads to an enhanced understanding of the structure growth of the Universe. In addition, a detailed statistical analysis of ${\mathcal{L}(\hat{\Theta}|data)}$, $\chi^2$, $\chi^2_\nu$, $\rm{AIC}$, $\Delta \rm{AIC}$, $\rm{BIC}$, and $\Delta \rm{BIC}$ has been carried out to validate the diffusive dark fluid model's performance.
\\
\\
We organize the rest of the manuscript as follows: In Section \ref{back}, we talk about the theory of the covariant thermodynamic description and come up with the field equations for the background Universe, focusing on the diffusive dark-fluid system. In this section, the full set of evolution equations is derived for the linear cosmological perturbation using the 1+3 covariant formalism. The evolution equations of the density contrast \(\delta(z)\) and the redshift space distortion $f\sigma_8(z)$ are presented to allow us to study structure formation. The constraining of cosmological parameters from MCMC simulations is done in Section \ref{resultanddiscussio} after the full sets of equations for the background and perturbation have been put in order. In this section, the numerical results of the work are broadly explained. This section also includes a statistical analysis of the work. Finally, in Sec. \ref{disc}, the conclusions are presented.
\section{Theoretical framework}\label{back}
Fundamentally, the standard $\Lambda$CDM cosmology arises as a solution to the Einstein field equations (EFEs), derived from the action:
\begin{equation}
S = \frac{c^4}{16\pi G}\int{d^{4}x \sqrt{-g} \left[ R + 2\left(L_m - \Lambda\right) \right]} \;,
\end{equation}
where $R$ is the Ricci scalar, $L_m$ is the matter Lagrangian density, and $\Lambda$ is the cosmological constant.
Then the corresponding EFEs are expressed as
\begin{equation}\label{efe}
G_{\mu\nu}+\Lambda g_{\mu\nu}=8\pi GT_{\mu\nu}\;,
\end{equation}
with the first (geometric) term is represented by the Einstein tensor, and $T_{\mu\nu}$ represents the total energy-momentum tensor (EMT) of matter fluid forms. Both $G_{\mu\nu}$ and $T_{\mu\nu}$ are covariantly conserved quantities. The EMT for perfect-fluid models is given by
\begin{equation}\label{emt}
T_{\mu\nu}=(\rho+p)u_\mu u_\nu+pg_{\mu\nu}\;,
\end{equation}
where $\rho$ and $p$ are the energy density and isotropic pressure of matter, respectively. Related by the barotropic equation of state (EoS), $p=w\rho$, for a constant EoS parameter $w$. The normalized vector  $u_\mu$ represents the four-velocity of fundamental observers moving with the fluid. The divergence-free EMT, $T^{\mu\nu}{}_{;\mu}=0$ leads to the fluid conservation equation
\begin{equation}
\dot{\rho}+3\frac{\dot{a}}{a}(1+w)\rho=0\;,
\end{equation}
where $a(t)$ is the cosmological scale factor whose evolution is given by the Friedmann equation
\begin{equation}\label{fe}
\frac{\dot{a}^2}{a^2}=\frac{8\pi G}{3}\rho+\frac{\Lambda}{3}-\frac{k}{a^2}\;,
\end{equation}
where $k$ is the normalized spatial curvature parameter with values $-1\;, 0\;,1$ depending on an open, flat, or closed spatial geometry, respectively. In our case, we assume a flat spatial geometry.  
\subsection{Background equations}
In a multi-component fluid system, it is usually assumed that the energy density of each perfect-fluid component evolves independently of the other fluids of the system:
\begin{equation}
\dot{\rho_i}+3\frac{\dot{a}}{a}(1+w_i)\rho_i=0\;,
\end{equation}
and in this case, the EMT in Eq.  \eqref{emt} is the algebraic sum of the EMTS of each fluid, so are the total energy density and total pressure terms of Eq. \eqref{fe} the algebraic sums of the individual components. However, suppose we relax this assumption due to the presence of diffusion between the constituent components of the fluid. In that case, the individual components do not obey the matter conservation equation, but the total fluid still does. For the $i$th component fluid, the new conservation equation  {given by \citep{haba2010energy,calogero2011kinetic,benisty2019unification} reads:
\begin{equation}\label{diffusivex}
T^{\mu\nu}_i{}_{;\mu}=N^\nu_i\;,
\end{equation}
where the current of the diffusion term for that fluid,  $N^\nu_i = \gamma_i u^\nu$,\; $\gamma_i$ represents the number density, and $u^\mu$: $u^\nu u_\nu =-1$ is the four-velocity of the fluids. In the works  \cite{haba2016dynamics,maity2019Universe, calogero2012cosmological,haba2010energy,calogero2011kinetic}, Eq. \eqref{diffusivex} is broadly discussed to express a particular form of interaction between DM and DE. 
Similar to \cite{haba2010energy}, we take into account the assumption that the dissipation results from a relativistic motion in a DE fluid. { The energy-momentum tensor for DE and DM components of the fluid, the conservation is given as 
\begin{eqnarray*}
    \nabla_\mu T^{\mu\nu}_{de} = -\nabla_\mu T^{\mu\nu}_{dm} = N^\nu  = \gamma u^\nu\;.
\end{eqnarray*}
The 00-component of the interaction term reads $N^0 = \gamma/a^3$ for a homogeneous Universe, where the full detail is presented in \citep{haba2016dynamics}}.  The non-conservation equation for $i^{th}$ fluid can be given by \cite{haba2016dynamics,maity2019Universe}:
\begin{equation}\label{generlconversation}
\dot{\rho_i}+3\frac{\dot{a}}{a}(1+w_i)\rho_i=\frac{\gamma_i}{a^3}\;, \quad \mbox{where}\; i = b,dm,de
\end{equation}
where $\gamma_i$  is a constant for two dark components that means  $\gamma_i$ stands for $\gamma_{dm}$ and $\gamma_{de}$ for the DM and DE respectively. The diffusive term vanishes for the baryonic matter $\gamma_{b} = 0$, fluid due to its non-interacting nature.} Integrating this equation \eqref{generlconversation} gives
\begin{equation}
\rho_i=a^{-3(1+w_i)}\left[\rho_{i0}+\gamma_i\int^{t}_{t_0}a^{3w_i}dt'\right] \label{density}\;,
\end{equation}
with $\rho_{i0}$ representing the present-day ($t=t_0$) value of the energy density of the $i$th fluid. Using a late-time $|t-t_0|\ll t_0$ expansion and expressing $a(t)=a_0\left[1-(t_0-t)H_0+\dots\right]$, we can write the last term of the above integrand as
\begin{eqnarray}\label{scalefactor}
\int^t_{t_0}a^{3w_i}dt&\approx&\int^t_{t_0}a^{3w_i}_0\left[1-(t_0-t')H_0+\dots\right]^{3w_i}dt'\nonumber\\
&&\approx \frac{a^{3w_i}_0}{(1+3w_i)H_0}\Bigg[\left(1-(t_0-t)H_0\right)^{1+3w_i}\nonumber\\&&
- \left(1+(t_0-t_0)H_0\right)^{1+3w_i}+\dots\bigg]\nonumber\\
&&  \approx \frac{a^{3w_i}_0}{(1+3w_i)H_0}\left[a^{1+3w_i}-1\right]\;.
\end{eqnarray}
In the last step, we normalized the scale factor to unity today: $a_0=1$. From Eqs. \eqref{density} and \eqref{scalefactor},  the energy density of each diffusive fluid component is given according to the relation given below:
\begin{equation}
\rho_i\approx a^{-3(1+w_i)}\left[\rho_{i0}+\frac{\gamma_i}{(1+3w_i)H_0}\left(a^{1+3w_i}-1\right)\right]\;.
\end{equation}
Assuming the well-known component of radiation is negligible\footnote{Since the contribution of radiation to the late-time cosmological expansion history is so minimal, we have safely neglected such a contribution in the analysis to be performed.}, {the energy density of dust-like matter $\rho_m = \rho_{b} +\rho_{dm}$ for the case of baryons and DM,  and the corresponding equation of state parameter, $w_m =  w_{b}=w_{dm} = 0$.  For the case of vacuum energy, we consider  $p_{de} = w_{de}\rho_{de}$, where the equation of state parameter is $w_{de} = -1$.} The above diffusive solution leads to:
\begin{eqnarray}
   \rho_{\rm m}=a^{-3}\left[\rho_{\rm m0}+\frac{\gamma_{\rm dm}}{H_0}\left(a-1\right)\right]\;,~\text{since}\;~\rho_{m0} = \rho_{\rm b0}+ \rho_{\rm dm0} 
\end{eqnarray}
{and}
\begin{eqnarray}
     \rho_{de}=\rho_{de 0}-\frac{\gamma_{de}}{2H_0}\left(a^{-2}-1\right)\;.
\end{eqnarray}
 It is worth mentioning at this point that the DM energy density $\rho_{dm}$ no longer scales like $a^{-3}$ since the extra items.
This equation can be rewritten in a way that resembles the DM scaling equation provided in \cite{naidoo2024dark}, according to which we can expect a change in the Integrated Sachs-Wolfe  (ISW) effect.   The modified Friedmann equation due to the presence of the diffusive dark fluid is given by
\begin{eqnarray}
H^2 = \frac{8\pi G}{3c^4} \Bigg[ \rho_{\rm m0}a^{-3} + \frac{\gamma_{\rm {dm}}}{H_0}\left(a-1\right)a^{-3} \nonumber\\+  \rho_{de0}  -\frac{\gamma_{de}}{2H_0}\left(a^{-2}-1\right)\Bigg].
\end{eqnarray}
Here we introduce the following dimensionless dynamical quantities:
\begin{eqnarray*}\label{Eq:h_z_1}
  && \Omega_{i}\equiv \frac{8\pi G}{3H^2_0}\rho_{i}\;, Q_{dm}\equiv\frac{8\pi G}{3H^3_0}\gamma_{dm}\;,Q_{de}\equiv\frac{8\pi G}{3H^3_0}\gamma_{de}\;,  h\equiv \frac{H(z)}{H_0}\;.  
\end{eqnarray*}
 Then, the normalised Hubble parameter yields\footnote{Note that the  matter density parameter is $\Omega_m = \Omega_{b} + \Omega_{dm}$. }
\begin{eqnarray}\label{normalizedhubbel}
  h^2(z) =\Omega_{m}(1+z)^3+\Omega_{de} -Q_{dm}(1+z)^2z  - \frac{1}{2} Q_{de}(z^2+2z)\;.  
\end{eqnarray}
In this work, we assume that the interaction exists only between the dark components and given the requirement $\sum \gamma_i = 0$ for total energy conservation. Hereafter, we set \(Q_{de} = -Q_{dm}\). 
 From Eq. \eqref{normalizedhubbel}, the simplified form of the deceleration parameter is yielded as
 \begin{equation} \label{deccx}
 q(z) =   -1 +   \frac{ (1+z)^2\left(3 \Omega_{m} (1+z) - 3 z Q_{dm}\right)}{2 \Big(\Omega_{m}(1+z)^3+\Omega_{de}
 -\frac{1}{3}Q_{dm} z^2 - Q_{dm} z^3  \Big)} \;. \end{equation}
 Then, the corresponding effective equation of state {parameter ($w_{eff}$) } for the diffusive dark fluid is obtained as
  \begin{eqnarray}\label{effective}
  w_{eff}(z)  = -\frac{1}{3} + \frac{2}{3}q(z)\;.
 \end{eqnarray}
Using Eq. \eqref{normalizedhubbel},  it is straightforward to compute  the cosmological distance measures such as the distance modulus $\mu(z)$, which is given by
 \begin{equation}\label{distancemodulus1}
 	\mu(z) =  25+5\log_{10}D_{L}(z)\;,
 \end{equation}
 where $D_L(z)$ is the luminosity distance. It is given by
 \begin{equation}
 D_{L}(z)=(1+z)300\bar{h}^{-1}\int_{0}^{z} \frac{c dz'}{h(z')'}\;.
 \end{equation}
 Here, $\bar{h} = H_0/100$, and $D_L(z)$ is measured in Mpc. The volume-averaged angular diameter distance reflects BAO measurements averaged over spherical distances. 
 \begin{equation}
  D_{V}(z)=\left[(1+z)^2 D^2_{A}(z)^{2}\frac{300\bar{h}^{-1}z}{h(z)}\right]^{\frac{1}{3}}~,    
 \end{equation}
 and the angular distance yields
 \begin{equation}
 D_{A}(z)=\frac{\bar{h}^{-1}}{(1+z)}\int_{0}^{z}\frac{dz'}{h(z')}\;.
 \end{equation}
 The sound horizon at the drag $r_d$ epoch is given by 
 \begin{equation}
 r_d=\int_{z_d}^{\infty}\frac{c_s(z)}{H(z)}dz~,   
 \end{equation}
 where $z_d$ is the redshift at drag epoch and $c_s(z)$ is the sound speed of the photon-baryon fluid.
\subsection{Perturbation equations}\label{perturbationsa}
In this section, we apply the 1+3 covariant gauge-invariant perturbation formalism to study the structure formation within the framework of diffusive dark-fluid cosmology. The $1+3$ covariant formalism decomposes spacetime into temporal and spatial components using the observer's four-velocity $u^a = \frac{dx^a}{d\tau}$, where $x^a$ denotes the coordinates and $\tau$ is the proper time. This approach helps to analyse crucial quantities such as the rate of fluid (volume) expansion  ($\Theta \equiv \tilde{\nabla}_a u^a$=3H), shear ($\sigma_{ab} = \sigma_{(ab)}$), and vorticity ($\omega_{ab} = \omega_{[ab]}$).  The Raychaudhuri equation governs the dynamics of expansion in a cosmological setting and plays a crucial role in understanding the formation of singularities; it is given by:
\begin{equation}
\dot{\Theta} = -\frac{1}{3}\Theta^2 - \sigma_{ab}\sigma^{ab} + \omega_{ab}\omega^{ab} - R_{ab}u^a u^b +\tilde{\nabla}^a\dot{u}_a\;,
\end{equation}
where, $R_{ab}$ is the Ricci curvature tensor, indicating the curvature of spacetime due to gravitational effects, and $R_{ab}u^a u^b = \frac{1}{2}\kappa(1+3w)\rho$ \cite{ellis1990density,dunsby1992cosmological}, where $\kappa \equiv \frac{8\pi G}{c^4}$.  In our case, we assume shear-free \(\sigma_{ab} =0\), and free rotational \(\omega_{ab} =0\) spacetime, and the Raychaudhuri equation is reduced to 
\begin{eqnarray}
    \dot{\Theta} = -\frac{1}{3}\Theta^2 - \frac{1}{2}\kappa(1+3w)\rho  +\tilde{\nabla}^a\dot{u}_a\;.
\end{eqnarray}
To simplify the structure formation, we assume that adiabatic perturbations where the equation of state parameter is constant $\dot{w} = 0$. To begin traditionally, we start by defining the gradient variables describing the fluctuations \cite{ellis1989covariant,ellis1990density,dunsby1992cosmological,ellis1999cosmological} in the individual energy densities $(i=m, \Lambda)$ and the volume expansion as 
\begin{equation}
D^i_{a}\equiv\frac{a\tilde{\nabla}_{a}\rho_i}{\rho_i}\;,\quad  Z_{a}\equiv a\tilde{\nabla}_{a}\Theta\;,
\end{equation}
where the tilde nabla operator $\tilde{\nabla}$ denotes the spatially projected covariant derivative.
These gradient variables evolve according to the following equations:
\begin{eqnarray}
 \dot{D}^i_a &=& - (1+w_i)Z_a +\Bigg[\frac{w_m(1+w_i)\Theta\rho_i}{(1+w_m)\rho_m}+\nonumber\\&&\left(\frac{w_m}{(1+w_m)\rho_m}+\frac{1}{\rho_i}\right)\gamma_ia^{-3}\Bigg] D^i_a  \;, \\
\dot{Z}_a  &=& - \frac{2}{3}\Theta Z_a - \left[\frac{1+3w_m}{2}+\frac{w_m}{(1+w_m)}\tilde{\nabla}^2\right]\rho_m D^m_a-  \rho_{\Lambda}D^\Lambda_a \;.~~~~~~~~
\end{eqnarray}
By setting the equation of state parameter $w_m = 0$ for DM and $w_{de} = -1$ for DE, the corresponding system of first-order evolution equations yields:
\begin{eqnarray}
   &&\dot{D}^m_a + Z_a -\frac{1}{\rho_m}\gamma_{dm}a^{-3} D^m_a = 0\;,\label{E1}\\
&&\dot{D}^{de}_a -\frac{1}{\rho_{de}}\gamma_{de} a^{-3} D^{de}_a = 0\;, \label{E2}\\
&&\dot{Z}_a +\frac{2}{3}\Theta Z_a+\frac{1}{2}\rho_{m}D^m_a+\rho_{de} D^\Lambda_a =0\;. 
\end{eqnarray}

We followed the following steps to get the evolution equations of density fluctuations from Eqs. \eqref{E1} and \eqref{E2}. In the first case,  we take into account the second-order time-dependent evolution equations from Eqs. \eqref{E1} and \eqref{E2}.  In the second case, we implement the scalar decomposition technique developed originally by the work in \cite{ellis1989covariant, dunsby1992cosmological,abebe2012covariant} to extract any scalar variable $Y$ using the relation $a\nabla_aY_b=Y_{ab}=\frac{1}{3}h_{ab}Y+\Sigma_{ab}^Y+Y_{[ab]}\;.$ Here $Y=a\nabla_a Y^a$, whereas $\Sigma^Y_{ab}=Y_{(ab)}-\frac{1}{3}h_{ab}Y$ and $Y_{[ab]}$ 
  represent the shear (distortion) and vorticity (rotation) of the density gradient field, respectively.  Then, we define the following scalar quantities as \cite{ ellis1989covariant, dunsby1992cosmological,abebe2012covariant,abebe2013large, ntahompagaze2018study, sahlu2020scalar,sami2021covariant,sahlu2025structure}
  \begin{eqnarray} \label{definenationDelta}
       {\Delta}_i = a\tilde{\nabla}^aD^i_a\;, \quad\mbox{and} \qquad Z = a\tilde{\nabla}^aZ_a
  \end{eqnarray}
   where $i$ stands for matter and DE components. 
    In the third step, we shall pay attention to solving the density contrast by resorting to the evolution equations \eqref{definenationDelta}, and we use the initial values of the system given by \cite{sahlu2020scalar, sahlu2024cosmology, sahlu2025structure,sahlu2025constraining}
\begin{eqnarray}\label{densitycontrastboth}
    \delta_i(z) = \frac{\Delta_i(z)}{\Delta_i(z_{in})}\;,
\end{eqnarray}
where the subscript $in$ refers to the initial value of $\Delta_i (z)$ at the given initial redshift $z_{in}$.
   Finally, we transformed the time-dependent to redshift space as presented in  \cite{sahlu2020scalar,sami2021covariant, sahlu2024cosmology,sahlu2025structure}   after following the scalar decomposition technique mentioned in the above second case\footnote{We consider a similar manner as Ref. in \cite{louis2019optimising}, which neglects the scale dependence, and the density contrast is rewritten as $\delta(k, z) =\delta({z}) $.}.
 Then the coupled second-order system of equations in redshift space becomes
 
\begin{eqnarray}
&&(1+z)^2\delta''_m+\left[(1+z)^2\frac{h'}{h}-(1+z)+\frac{Q_{dm}}{\Omega_m}\frac{1}{h}\right]\delta'_m+ \nonumber\\&& \frac{1}{h^2}\bigg[\frac{Q^2_{dm}}{\Omega^2_m}(1+z)^6-\frac{2Q_{dm}}{\Omega_m}(1+z)^3h  -\frac{3\Omega_m}{2} (1+z)^3\bigg]\delta_m \nonumber\\&&  -3\Omega_{de}\delta_{de}=0\;,\label{matter}\\&&
(1+z)^2\delta''_{de}+\left[(1+z)^2\frac{h'}{h}+(1+z)-\frac{Q_{dm}}{\Omega_{de}}(1+z)^4\frac{1}{h}\right]\delta'_{de} \nonumber\\&&-\frac{1}{h^2}\bigg[\frac{Q^2_{dm}}{\Omega^2_{de}}(1+z)^6 -3\frac{Q_{dm}}{\Omega_{de}}(1+z)^3h\bigg]\delta_{de}=0\;,\label{cosmologicalconstant} 
\end{eqnarray}
where $'$ and $''$ represent the first- and second-order derivatives with respect to the redshift, $z$.  {This coupled system of equations  \eqref{matter} and \eqref{cosmologicalconstant}  indicates that the cosmological perturbation of DE is not identically zero, and further study has been found in the work \citep{he2009effects}, where the influence of non-vanishing DE perturbations is considered.  The numerical results of these coupled system equations are presented later in Section \ref{perturbationdynamics}, and the result is strongly favored for the $\delta_m\gg\delta_{de}$. Hereafter, we extended further investigations of the large-scale structure by considering the assumption \(\delta_m(z)\gg\delta_{de}(z)\), that the DM component makes a significant contribution to structure formation. Without DM, the structure formation would not have enough time to grow.}   By admitting this assumption, we have a closed system of evolution equations from Eq. \eqref{matter}, and it is given as 
  \begin{eqnarray}
&&(1+z)^2\delta''_m+\left[(1+z)^2\frac{h'}{h}-(1+z)+\frac{Q_{dm}}{\Omega_m}\frac{1}{h}\right]\delta'_m\\&&+\frac{1}{h^2}\bigg[\frac{Q^2_{dm}}{\Omega^2_m}(1+z)^6-\frac{2Q_{dm}}{\Omega_m}(1+z)^3h-\frac{3\Omega_m}{2} (1+z)^3\bigg]\delta_m=0\;.~~~~\label{densityevolution}\;\nonumber
 \end{eqnarray}   \label{rry
 }
\\
\\
The diffusive dark-fluid model treats the dark sector as a single fluid with energy exchange between DM and DE, influencing the evolution of cosmological perturbations and potentially altering the growth of cosmic structures compared to the standard \(\Lambda\)CDM model \cite{naidoo2024dark}. The diffusive model's damping effect affects matter distribution and leaves observable imprints on the CMB, providing an alternative explanation for the Universe's accelerated expansion that can be tested through galaxy surveys and observations of structure growth.
The normalised density contrast \( \delta_m(z)\), plays a key role in the formation of cosmic structures. It starts small, growing through gravitational instability to form galaxies and clusters. Afterward, we have considered the growth factor $D(z)$  that represents the ratio of the amplitude of \( \delta_m(z)\) in redshift $z$ compared to an initial value $\delta_m(z=0)$ becomes
\begin{eqnarray}\label{ss}
     D(z) = \frac{\delta_m(z)}{\delta_m(z=0)}\;.
 \end{eqnarray}
 It is often normalised to $\delta(z =0) = 1$ and is governed by a differential equation that involves the Hubble parameter the density of matter. This factor shows how initial density perturbations grow over time due to gravity, influencing the formation of large-scale structures. In a DE-dominated Universe, the growth slows due to accelerated expansion. The growth factor is crucial for modeling galaxy formation and comparing theoretical predictions with observations, such as those from the CMB and galaxy surveys \cite{springel2006large}. Additionally, the growth rate $f(z)$, related to $D(z)$, measures structure growth and is used in observational probes such as redshift-space distortions ${f}\sigma_8$. From Eq. \eqref{ss}, the growth rate $f(z)$ is  yields as 
\begin{equation}\label{growth11x}
    f 
    \equiv       
    \frac{{\rm d}\ln{{{D}}}}{{\rm d}\ln{a}} = -(1+z)\frac{\delta'_m(z)}{\delta_m(z)}    
    \;.
\end{equation}
Therefore, the growth factor is fundamental to understanding the dynamical evolution of the structures of the universe.
Thus, substituting the definition of \eqref{growth11x} into the second-order evolution equation \eqref{densityevolution}, the growth rate is governed by
\begin{eqnarray}\label{growthrate}
    &&(1+z)f' =  f^2 - \left((1+z)\frac{h'}{h}- 2+\frac{Q_{dm}}{\Omega_m}\frac{1}{h(1+z)}\right)f \\&&-\frac{1}{h^2}\bigg(\frac{Q^2_{dm}}{\Omega^2_m}(1+z)^6-\frac{2Q_{dm}}{\Omega_m}(1+z)^3h-\frac{3\Omega_m}{2} (1+z)^3\bigg)\;.\label{growth1}\nonumber
\end{eqnarray} 
For the case of $Q_{dm} = 0$, the evolution equation of the density fluctuation for $\Lambda$CDM is recovered.
A combination of the linear growth rate $f(z)$ with the root-mean-square normalization of the matter power spectrum $\sigma_8$ within the radius sphere $8h^{-1}$Mpc, yields the redshift-space distortion $f\sigma_8$  \cite{hamilton1998linear} as
\begin{eqnarray}\label{growth11}
  f\sigma_8(z)  = -(1+z)\sigma_8\frac{\delta'_m(z)}{ \delta_m(z)}\;.
 \end{eqnarray}
Note that all the above background and perturbation evolution equations are reduced to the $\Lambda$CDM model for the closure of vanishing interaction term, $Q_{dm} = 0$. In the subsequent section, as presented in \ref{resultanddiscussiox}, we emphasize the matter clustering $S_8$ that is given by
$$S_8 = \sigma_8\sqrt{\frac{\Omega_m}{0.3}}\;,$$ to explain the viability of the diffusive model to explain the cosmic structure growth and matter distribution in the Universe. 
\begin{figure*}
\includegraphics[scale=0.55]{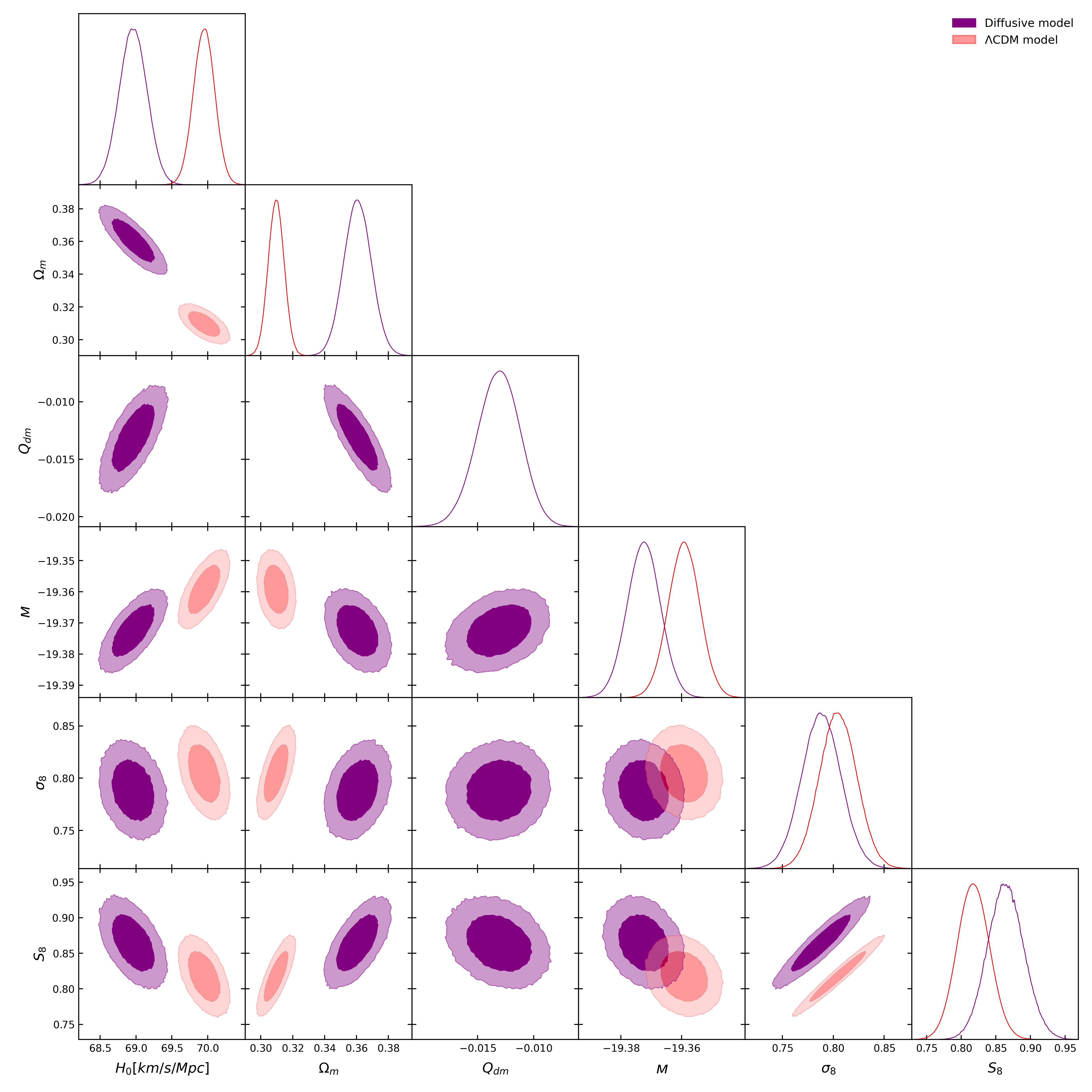}
\caption{Posterior distributions of the parameters  using   \textit{ PPS  + CC + DESY5 + RSD + f} for  both models.}\label{corner1}
\end{figure*}
\section{Results and discussion}\label{resultanddiscussio}
Using the recent cosmological measurements, this section gives a detailed analysis of the constraining parameters,  the comparison of $H_0$ (in km/s/Mpc) and $S_8$ values with different cosmological surveys, the numerical result of background cosmological parameters, structure growth, and statistical analysis that show the implications of the diffusive model against the $\Lambda$CDM model. 
\begin{table*}
\caption{Marginalized 68\% and 95\% confidence level (C.L) limits of the parameters in the dataset combinations
for the diffusive model and the $\Lambda$CDM model.}
{
\begin{tabular}{|lc|c|c|c|}
\hline
Parameter &~~~~~~ C.L~~~~~~~~~~~~&  \textit{ PPS  + CC + DESY5}  &  \textit{DESI DR2 BAO + CC} & \textit{DESI DR2 BAO + CC + DESY5} \\
 & &  \textit{+ RSD + f}  & \textit{+ Union3 + RSD + f} &\textit{Union3 + RSD + f}  \\
\hline
 Diffusive model&&&&\\
 &&&&\\
$H_0$ (km/s/Mpc) & 68\% & $ 68.959^{+0.193}_{-0.193}$ & $ 69.469^{+0.169}_{-0.169}$&$ 69.465^{+1.392}_{-1.385}$
\\
&95\% & $ 68.959^{+0.378}_{-0.377}$ &$ 69.469^{+0.333}_{-0.333}$&
 $ 69.465^{+2.761}_{-2.730}$
\\
&best-fit &  68.959 & 69.469& 69.465 \\
$\Omega_m$ & 68\% & $ 0.361^{+0.009}_{-0.009}$ &$ 0.331^{+0.007}_{-0.007}$&  $ 0.309^{+0.008}_{-0.007}$\\
&95\% &$ 0.361^{+0.017}_{-0.017}$ & $ 0.331^{+0.013}_{-0.013}$ &
 $ 0.309^{+0.015}_{-0.014}$ \\
&best-fit & 0.361 &0.331&0.309\\
$r_d$ & 68\%  & -- & $ 142.239^{+0.649}_{-0.644}$& $ 144.726^{+2.920}_{-2.808}$\\
&95\% & -- & $ 142.239^{+1.278}_{-1.269}$&$ 144.726^{+5.873}_{-5.463}$
\\
&best-fit  & -- & 142.239 &144.726\\
$Q_{dm}$ & 68\% &  $ -0.013^{+0.002}_{-0.002}$ & $ -0.007^{+0.001}_{-0.001}$ & $ -0.004^{+0.001}_{-0.001}$
\\
&95\% &$ -0.013^{+0.004}_{-0.004}$& $ -0.007^{+0.003}_{-0.003}$& $ -0.004^{+0.003}_{-0.003}$
 \\
&best-fit &$-0.013$ &$-0.007 $&$-0.004$\\
$M$ & 68\% &  $ -19.373^{+0.005}_{-0.005}$& --&-- \\
&95\% & $ -19.373^{+0.011}_{-0.011}$ &--& --\\
&best-fit & $-19.373$ & --&--\\
$\sigma_8$ & 68\% & $ 0.788^{+0.019}_{-0.019}$ &  $ 0.771^{+0.019}_{-0.018}$ &  $ 0.756^{+0.018}_{-0.018}$
 \\
&95\% &$ 0.788^{+0.038}_{-0.037}$ & $ 0.771^{+0.037}_{-0.036}$ & $ 0.756^{+0.036}_{-0.035}$ \\
 & best-fit&  0.788  &0.767 &\\
$S_8$ & 68\% &  $ 0.864^{+0.027}_{-0.026}$ &$ 0.809^{+0.023}_{-0.022}$& $ 0.766^{+0.023}_{-0.023}$
 \\
&95\% &$ 0.864^{+0.053}_{-0.050}$ & $ 0.809^{+0.046}_{-0.044}$ & $ 0.766^{+0.046}_{-0.044}$ \\
best-fit & &0.864 & 0.809&0.766 \\
\hline
 $\Lambda$CDM model&&&&\\
$H_0$ & 68\% &$ 69.950^{+0.146}_{-0.147}$& $ 69.968^{+0.143}_{-0.143}$&$ 70.101^{+1.386}_{-1.372}$
 \\
&95\% & $ 69.950^{+0.287}_{-0.288}$& $ 69.968^{+0.282}_{-0.281}$ & $ 70.101^{+2.753}_{-2.700}$\\
&best-fit & 69.950 & 69.968 &70.101\\
$\Omega_m$ & 68\% &  $ 0.310^{+0.005}_{-0.005}$&$ 0.306^{+0.005}_{-0.004}$& $ 0.293^{+0.005}_{-0.005}$
 \\
&95\% &  $ 0.310^{+0.010}_{-0.010}$ &$ 0.306^{+0.009}_{-0.009}$ &  $ 0.293^{+0.009}_{-0.009}$
 \\
&best-fit & 0.310 &0.306 &0.293\\
$r_d$ & 68\%  & --& $ 144.197^{+0.560}_{-0.560}$ & $ 145.369^{+2.923}_{-2.831}$ \\
&95\% & -- & $ 144.197^{+1.109}_{-1.099}$ & $ 145.369^{+5.861}_{-5.504}$\\
&best-fit  & -- & 144.197 &  145.369\\

$M$ & 68\% & $ -19.359^{+0.005}_{-0.005}$ & --& \\
&95\% &$-19.359^{+0.010}_{-0.010}$ &  -- &\\
&best-fit & $-19.359$& --  &\\
$\sigma_8$ & 68\% &$0.805^{+0.018}_{-0.018}$ & $ 0.796^{+0.018}_{-0.017}$ & $ 0.765^{+0.018}_{-0.017}$
\\
&95\% & $ 0.805^{+0.036}_{-0.035}$ & $ 0.796^{+0.035}_{-0.034}$ & $ 0.765^{+0.035}_{-0.034}$\\
 & best-fit&  0.805 &0.796 &0.765\\
$S_8$ & 68\% & $ 0.818^{+0.023}_{-0.023}$ &   $ 0.804^{+0.022}_{-0.021}$ & $ 0.756^{+0.022}_{-0.021}$\\
&95\% & $ 0.818^{+0.046}_{-0.045}$ & $ 0.804^{+0.044}_{-0.042}$
& $ 0.756^{+0.044}_{-0.042}$\\
 & best-fit & 0.818&0.804 &0.756\\
\hline
\end{tabular}
\label{table-bestfit}
}
\end{table*}
\subsection{Constraining parameters}\label{resultanddiscussiox}
We consider the Python libraries, including EMCEE \cite{foreman2013emcee, hough2020viability} and GetDist \cite{lewis2019getdist}, to constrain the values of the parameters using the mentioned in the below cosmological measurements. We have considered the following recent cosmological measurements to constrain the model parameters,

\begin{enumerate}
    \item \textit{BAO:} The BAO distance and the correlation measurements released data 2 (DR2) \cite{andrade2025validation, abdul2025desi} from the Dark Energy Spectroscopic Instrument (DESI) Survey have been considered. The measurements include data for the isotropic BAO measurements of $D_V(z)/r_d$, where $D_V(z)$ and $r_d$ are the spherically averaged volume distance and sound horizon at baryon drag, respectively. And anisotropic BAO measurements of $D_M(z)/r_d$ and $D_H(z)/r_d$, where $D_M(z)$ and $D_H(z)$ are the co-moving angular diameter distance and the Hubble distance, respectively, and the correlations between the isotropic and anisotropic BAO measurements.  Hereafter, we refer to this dataset as \textit{DESI DR2 BAO }.
  
    \item \textit{Supernovae Type Ia (SNIa)} dataset compilations we have considered the following, namely: I) \textit{PPS:} we use the SNIa distance moduli measurements from the Pantheon+ sample \cite{brout2022pantheon+}, which consists of 1701 light curves of 1550 distinct SNIa ranging in the redshift interval $z \in [0.001, 2.26]$,    II) \textit{DESY5} data \cite{collaboration2024dark} which is a photometrically-classified SNIa with redshifts in the range $0.1 < z < 1.13$, complemented by 194 historical low-redshift SNe Ia (which are also present in the PPS sample) spanning $0.025 < z < 0.1$, iii) \textit{Union3} we have consider the up-to-date Union compilation of 2087 cosmologically useful SNIa from 24 datasets \cite{rubin2023union}.  
\begin{figure*}
\includegraphics[scale=0.55]{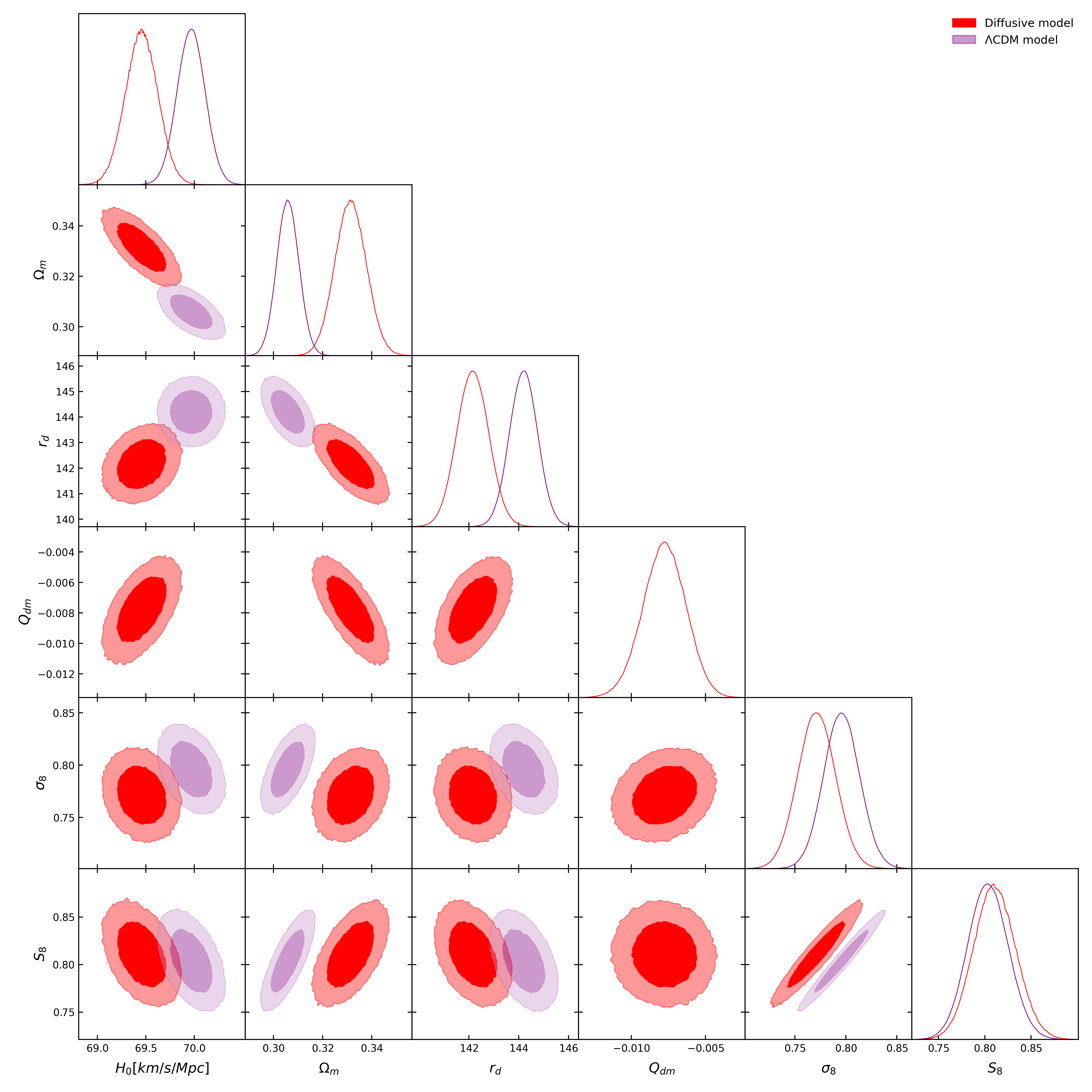}
\caption{Posterior distributions of the parameters  using  \textit{DESI DR2 BAO + CC + DESY5 + RSD + f} for  both models.}\label{corner2}
\end{figure*}
    \item \textit{Cosmic Chronometers}: 
      We analyse the Hubble parameter $H(z)$ measurements with observational Hubble parameter data.  This comprises 31 data points derived from the relative ages of massive, early-time, passively evolving galaxies, known as cosmic chronometers (CC).  We refer to this dataset as {CC}.
We calculate the minimum $\chi^2$ with cosmic chronometer covariance in combination with statistical and systematic effects as presented in chronometers (CC) \cite{moresco2020setting,qi2023model}
\begin{equation*}
     \chi^2_{CC} = \left(H_{\text{theo}}(z)-H_{\text{ob}}(z_i)\right)^{T}C^{-1}\left(H_{\text{theo}}(z)-H_{\text{ob}}(z_i)\right),
\end{equation*}
where $H_{theo}(z)$ represents the theoretical model of the Hubble parameter and$H_{ob}(z_i)$ the Hubble parameter measurement.
    \item  \textit{Large scale structure}: We also incorporate redshift-space distortion data and the growth rate, labeled {RSD},  from the VIMOS Public Extragalactic Redshift Survey (VIPERS) and SDSS collaborations. i) A total of 66 data points of the measurements of redshift-space distortion for ${f}\sigma_8$ have been collected and summarized in the works of  \cite{kazantzidis2018evolution,skara2020tension}, covering the redshift interval $0.001 \leq z \leq 1.944$. 
ii)14 growth rate \textit{f} data points within the redshift range $0.001 \leq z \leq 1.4$ \cite{woodfinden2022measurements}.
The resulting $\chi^2$ for redshift-space distortion  and growth rate are expressed as:
 \begin{equation*}
     \chi^2_{\text{RSD}} = \left(f\sigma_{8, \text{theo}}-f\sigma_{8,\text{ob}}(z_i)\right)^{T}C^{-1}\left(f\sigma_{8, \text{theo}}-f\sigma_{8,\text{ob}}(z_i)\right)\;,
\end{equation*}
{and} 
\begin{eqnarray*}
\chi^2_{f} = \left(f_{\text{theo}}-f_{\text{ob}}(z_i)\right)^{T}C^{-1}\left(f_{\text{theo}}-f_{\text{ob}}(z_i)\right)
\end{eqnarray*}
receptively. 
\item \textit{Joint datasets}: The combined data analysis improves the precision and constraints on the models we employ, allowing for a more complete picture of the Universe.  By considering the inconsistency of the \textit{PPS} \&  DESI $BAO$ datasets as presented in the  \cite{afroz2025hint} disclosed through a violation of the distance duality relation, we have considered a joint analysis
\begin{enumerate}
    \item  \textit{ PPS  + CC + DESY5 + RSD + f}
\item \textit{DESI DR2 BAO + CC + Union3 + RSD + f} and 
\item \textit{DESI DR2 BAO + CC +DESY5+ Union3 + RSD  f}
\end{enumerate}
for both theoretical models. The corresponding $\chi^2$ is given by
\begin{eqnarray*}
   \chi^2_{tot} = \sum_k{\chi^2_k}\;,  
\end{eqnarray*}
where $k$ stands for \textit{DES BAO, PPS, CC, DESY5, Union3, RSD}, and \textit{f}
\end{enumerate}

\begin{figure*}
\includegraphics[scale=0.55]{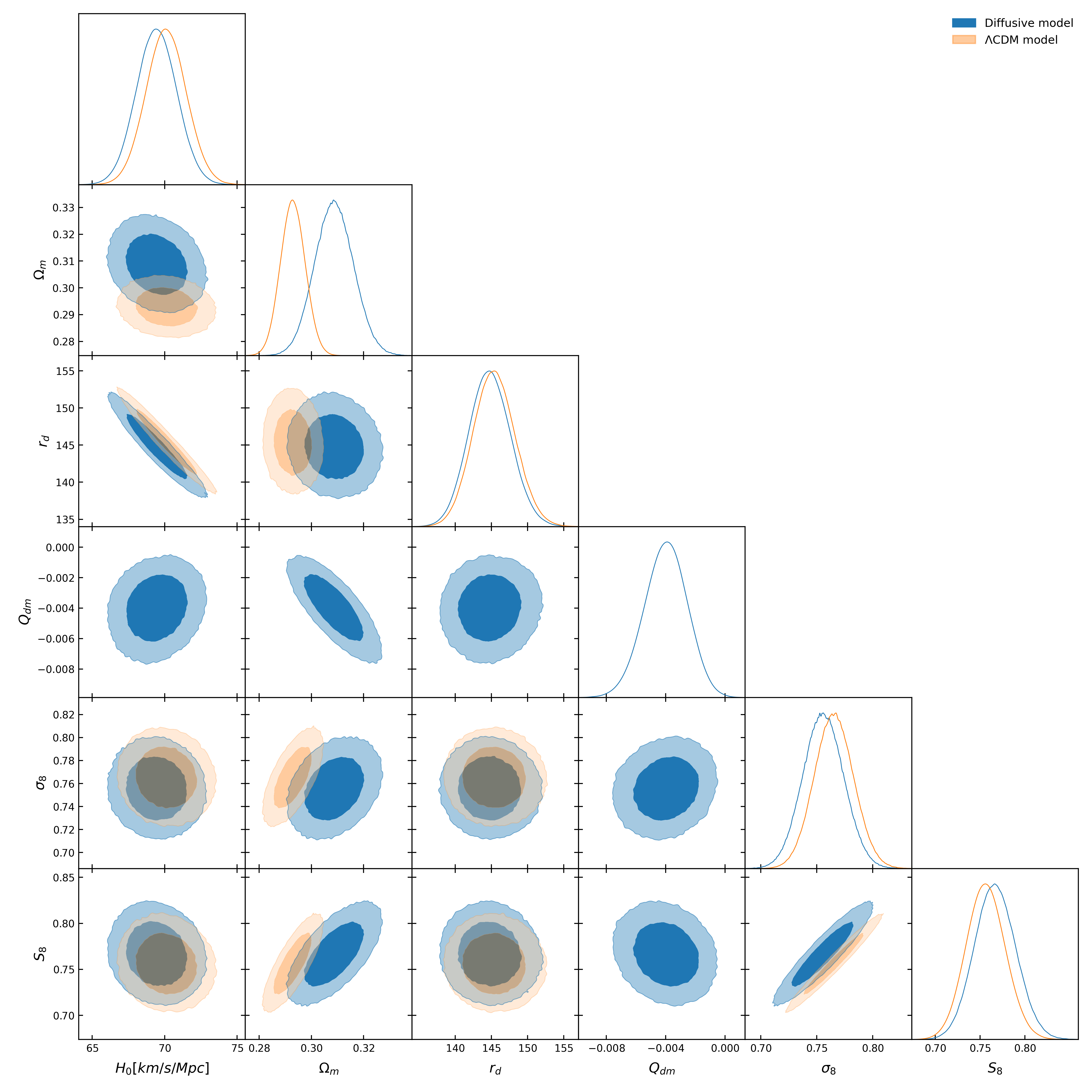}
\caption{Posterior distributions of the parameters  using  \textit{DESI DR2 BAO + CC + Union3 + RSD + f}  for  both models.}\label{corner3}
\end{figure*}

In the present work, we have emphasized the above two joint datasets to constrain the cosmological parameters $h,\Omega_{m}$, $Q_{dm}$, $r_d$, $M$, and $\sigma_8$, and for detailed statistical analysis, see Table \ref{table-bestfit}. The values of our constrained parameters are highly sensitive to the prior on $Q_{dm}$, and we have set the following priors: \(\Omega_m = [0.1, 1.0]\), \(H_0 = [0.0, 1.0]\), \(r_d = [100, 200], Q_{dm} = [-0.05, 0.05],~\text{and}~M = [-15.00, 22.00]\). We present the posterior distributions of these parameters in Figs.  \ref{corner1},  \ref{corner2} and \ref{corner3} for  the $\Lambda$CDM and  diffusive model respectively.  The parameter values are summarized in the marginalized 68\% and 95\% confidence limits in Table \ref{table-bestfit} for joint datasets, \textit{ PPS  + CC + DESY5 + RSD + f}, \textit{ DESI DR2 BAO  + CC +DESY5 + RSD + f}, and \textit{ DESI DR2 BAO  + CC + Union3 + RSD + f}. We notice that the value of the interaction term is negative, see the best-fit values in Table \ref{table-bestfit}, \(Q_{dm} = -0.013, \, -0.007, \, -0.004\) at 68\%  and  95\% C.Ls for the considered joint datasets.  This indicates a case where energy flows from DM to DE, driving the Universe's accelerating expansion, as supported by recent observational evidence \citep{van2025compartmentalization,benisty2019unification,silva2025new}.  In the subsequent sections, we shall choose 95\% of C.L of the values of the interacting term \(Q_{dm} = -0.013^{+0.004}_{-0.004} = -0.017,-0.013, -0.009\)  as listed in Table \ref{table-bestfit}  to show our numerical results. 
\begin{figure*}
    \includegraphics[width=0.45\linewidth]{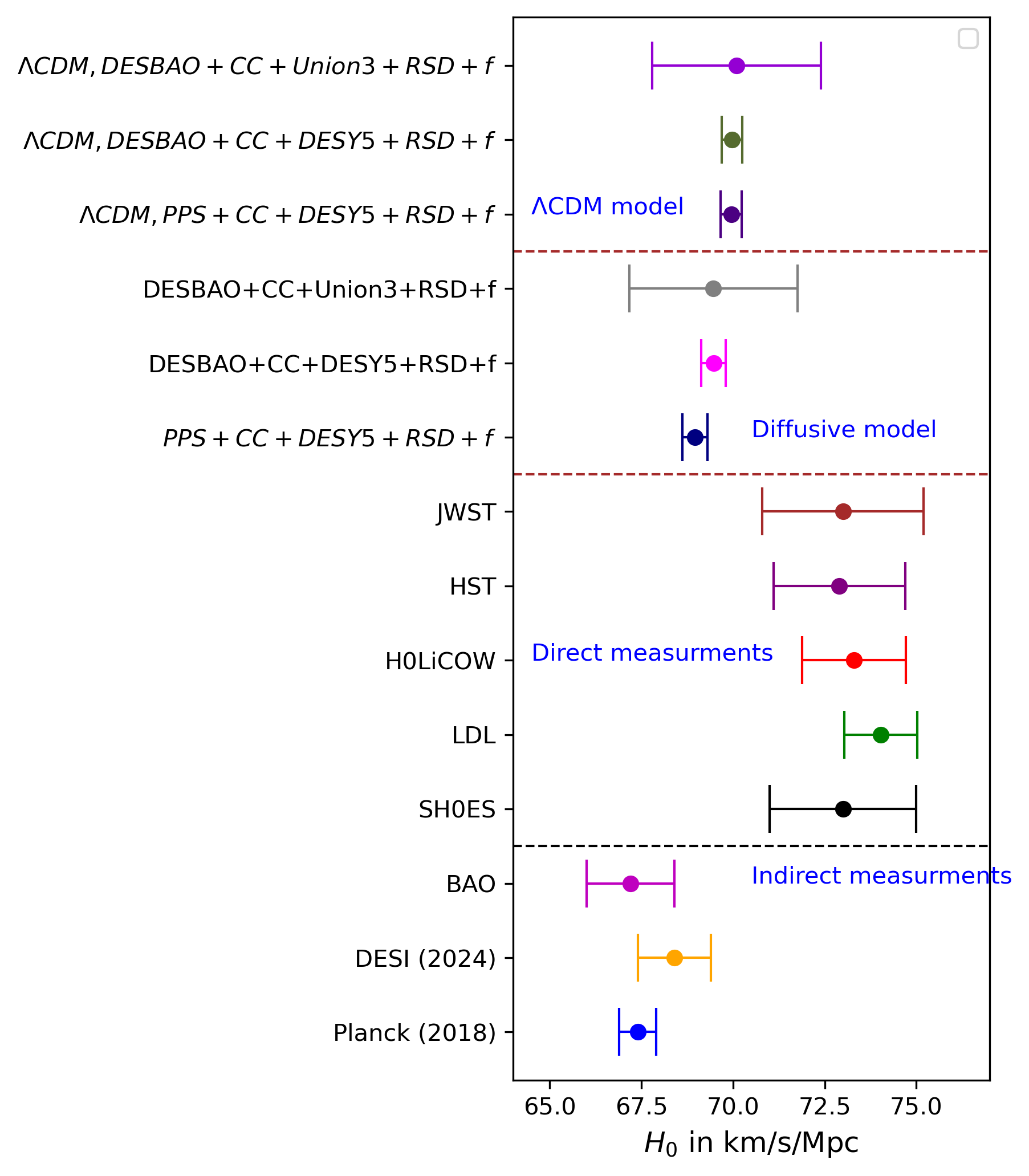}
    \qquad
    \includegraphics[width = 0.45\linewidth]{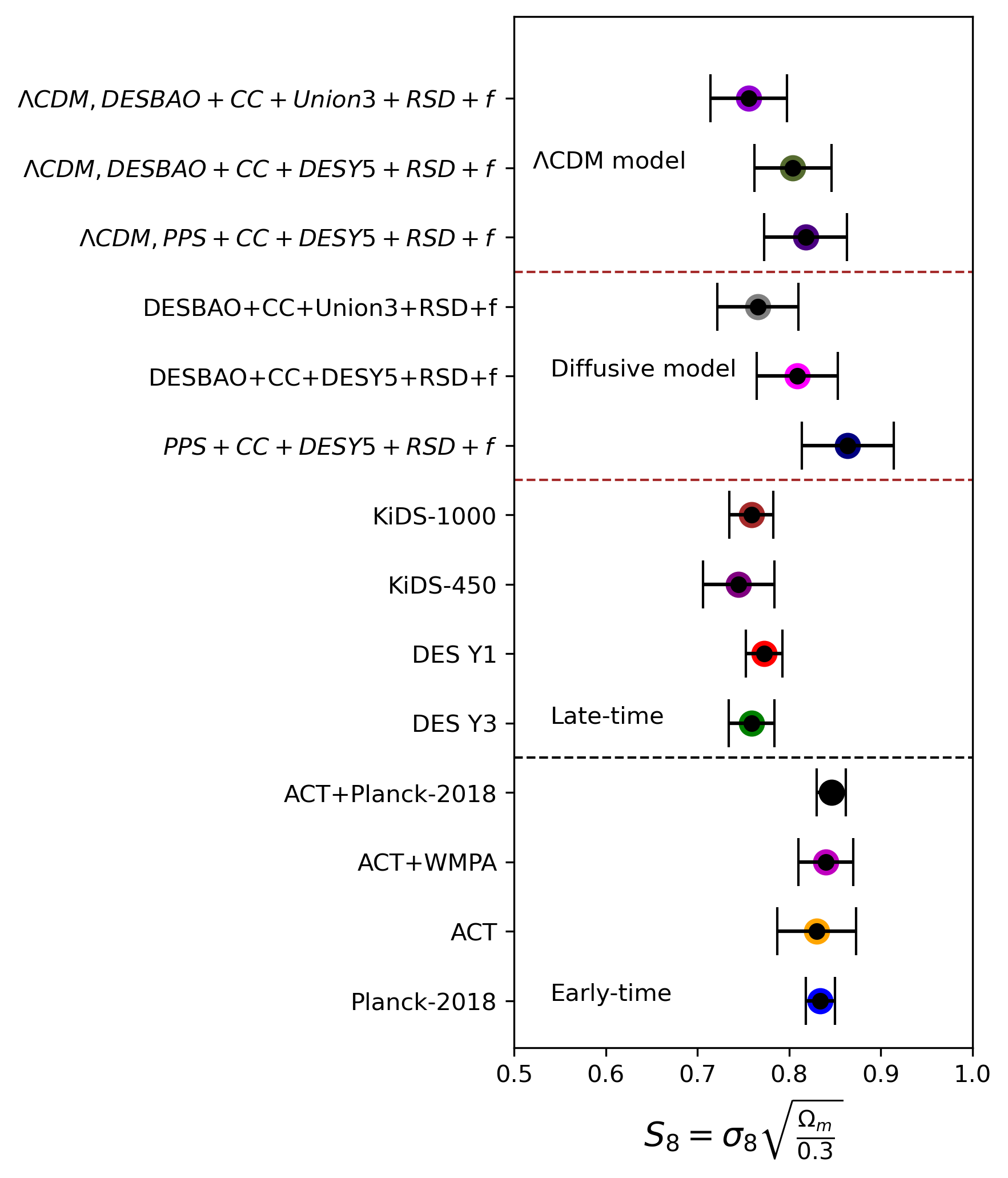}
    \caption{The comparison of $\Lambda$CDM and diffusive model $H_0$ values (in km/s/Mpc) with early and direct measurements (left panel), \(S_8\) values between the \(\Lambda\)CDM and diffusive models alongside various late-time and early-time measurements (right panel).}
    \label{fig:enter-labelH0}
 \end{figure*}
 \begin{figure*}
    \includegraphics[width=0.9\linewidth]{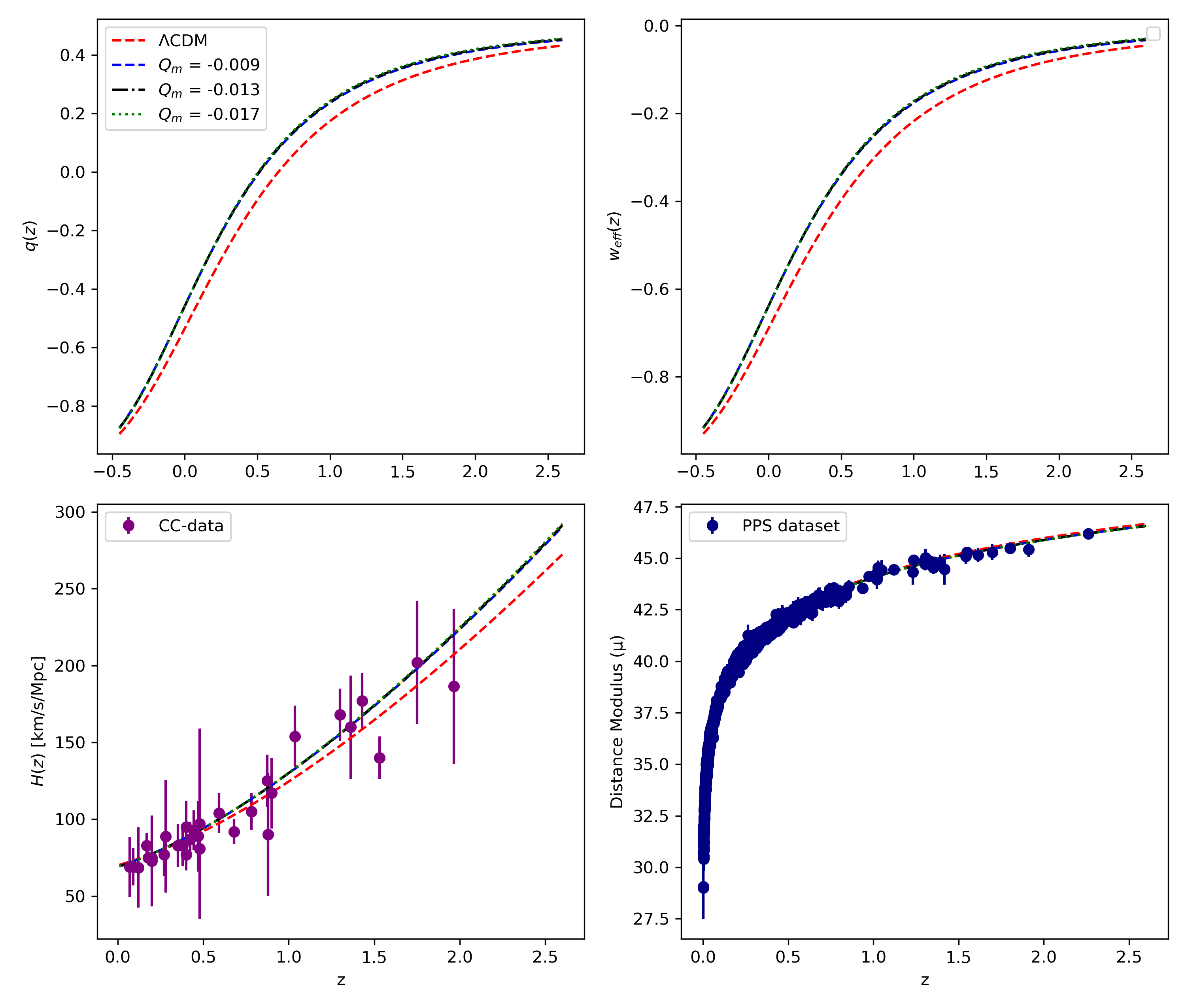}
    \caption{The diagrams for multiple cosmological parameters in both models: diffusive dark-fluid and \(\Lambda\)CDM models are highlighted in this Fig. i). With slight variations for the diffusive model at higher redshifts, the top-left panel deceleration parameter \( q(z) \) depicts the change from deceleration to acceleration. ii) The effective equation of state parameter \( w_{\text{eff}}(z) \) in both models is shown in the upper-right panel. iii) The bottom-left panel shows that the CC data for both models fit the Hubble parameter \( H(z) \). iv) The bottom-right panel shows the distance modulus $\mu(z)$, which both models closely match the data with PPS. For illustrative purpose we  use $Q_{dm} = -0.017, -0.013,-0.009$, $H_0 = 68.959$ in km/s/Mpc, $\Omega_m = 0.361$   for diffusive model and $\Omega_m = 0.310$ and $H_0 = 69.950$ in km/s/Mpc for $\Lambda$CDM model from  joint analysis of \textit{ PPS  + CC + DESY5 + RSD + f} presented in Table \ref{table-bestfit}. }
    \label{fig:enter-labelback}
\end{figure*}
\\
\\
{From the constrained parameters presented in the Table. \ref{table-bestfit} and  Figs. \ref{corner1}, the diffusive model has slightly lower values of $H_0$ (km/s/Mpc) and $\sigma_8$ but higher values of $\Omega_m$ introducing \(\Omega_m\) tension compared with $\Lambda$CDM model which leads to the slightly higher deviation of the values $S_8$ between both models. This deviation is significantly noticed for \textit{ PPS  + CC + DESY5 + RSD + f} datasets, where the best fit values of $\Omega_m = 0.361$ for the diffusive model, which leads to $S_8$ tension. For instance, at 95\% confidence level the value of the density parameter \(\Omega_m = 0.361^{+0.017}_{-0.017}\)  which is higher in the diffusive model than the \(0.310^{+0.010}_{-0.010}\)  in the \(\Lambda\)CDM model (see Table. \ref{table-bestfit} ).  The deference is by $2.6\sigma$ which is a significant deviations. Similarly, this deviation is \(2.3\sigma\) when comparing the diffusive model (using the same \(\Omega_m\) value) to Planck's measured \(\Omega_m = 0.315 \pm 0.01\) \citep{aghanim2020planck}}. Notably, these tensions are significantly alleviated when considering data combinations involving DESI DR2 with CC, DESY5, and Union3, as presented in Table \ref{table-bestfit} and Figs. \ref{corner2} and \ref{corner3}.
\\
\\Indeed, the increased precision in Hubble parameter (\(H_0\) km/s/Mpc) measurements indicates a disparity between indirect and direct measurements in estimating the present-day expansion rate of the Universe, known as the Hubble tension \cite{riess20113, riess2019large,riess2022comprehensive,aghanim2020planck,di2021combined,verde2019tensions,wong2020h0licow,adame2025desi}. Besides,  the discrepancy in the inferred value of the matter clustering parameter has been reported based on the observations of Planck-2018, KiDS-1000 \cite{hildebrandt2017kids}, KiDS-450) \cite{asgari2021kids} DES Y1 \cite{abbott2018dark}, DES Y3 \cite {amon2022dark} and $f\sigma_8(z)$ measurements from RSD mentioned a few. Unlike the discrepancy, measurements of  $H_0$ (km/s/Mpc)  and $S_8$ remain unresolved, providing new cosmological insights beyond the $\Lambda$CDM model \cite{di2021cosmology}. In this section, we compare the $H_0$ (km/s/Mpc) and $S_8$ values of the diffusive model to those of the $\Lambda$CDM model alongside various measurements, indicating that the diffusive model gained the capability to mitigate cosmological tensions.  To do that, 
{we first shall take into consideration the $H_0$ measurements with $i)$ the indirect measurements, including Planck 2018  (\(H_0 = 67.4 \pm 0.5\) km/s/Mpc) \citep{aghanim2020planck}, DESI-2024 (\(H_0 = 68.52\pm 0.62\) km/s/Mpc) \citep{adame2025desi}; and $ii)$ the direct measurements} derived from direct observations of the local Universe, such as Supernovae and $H_0$ for SH0ES (\(H_0 = 74.03 \pm 1.42\)) \citep{riess2019large}, H0LiCOW  (\(H_0 = 73.3 \pm 1.8\)) \citep{wong2020h0licow}, and HST (\(H_0 = 73.8 \pm 2.4\) km/s/Mpc)  \citep{riess20113}. For instance, the deviation between the SNIa (SH0ES) and Planck 2018 measurements of $H_0$ is $5.28\sigma$. From the results in Table~\ref{table-bestfit}, the diffusive model's $H_0=100h$ values (the 95\% C.L considered) for the datasets \textit{PPS + CC + DESY5 + RSD + $f$}, \textit{DESI DR2 BAO + CC + DESY5 + RSD + $f$}, and \textit{DESI DR2 BAO + CC + Union3 + RSD + $f$} differ from the Planck 2018 measurement by $2.49\sigma$, $3.44\sigma$, and $0.74\sigma$, respectively. Similarly, the diffusive model differ from SNIa (SH0ES) by  $-3.45\sigma$, $-3.12\sigma$, and $-1.48\sigma$ for the same datasets respectively. \textit{Note that, the negative sign in the sigma difference represents the measured value is higher than the hypothetical model's values.} The corresponding differences between the $\Lambda$CDM $H_0$ values (from the same table) and the Planck measurement increase to $4.42\sigma$, $4.47\sigma$, and $0.94\sigma$, respectively. In the same manner, differences between the $\Lambda$CDM $H_0$ values and SNIa (SH0ES) is $-2.81\sigma$, $-2.80\sigma$, and $-1.28\sigma$.   All these values of $H_0$ in km/s/Mpc are presented in Fig. \ref{fig:enter-labelH0} (left panel) together with the direct and indirect measurements.     From this Figure and above mentioned sigma compression, we notice that the diffusive model is favored to indirect measurements while the $\Lambda$CDM model is favored by direct measurements. Overall, the diffusive model shows smaller sigma differences compared with the $5.28\sigma$ tension and thus partially relax the tension, which requires further investigation.

   
 Secondly, we pay attention to matter clustering $S_8$, a critical cosmological measure that evaluates the growth of structure in the Universe. The observational tension between CMB and large-scale structure surveys may hint at novel physics beyond the $\Lambda$CDM model. We constrain the $S_8$ results in Fig. \ref{fig:enter-labelH0} (right-panel) for both theoretical models
using the above-mentioned datasets. 
 We have made a comparison of different \(S_8\) measurements findings which are highlighted in Fig. \ref{fig:enter-labelH0} (right panel) resorting to $i)$ late-time measurements, namely, KiDS-1000 ($S_8 = 0.759^{+0.024}_{-0.021}$) \cite{asgari2021kids}, KiDS-450 ($S_8 = 0.745^{+0.039}_{-0.039}$) \cite{hildebrandt2017kids}, DES Y1 ($S_8 = 0.759^{+0.025}_{-0.023}$) \cite{abbott2018dark}, DES Y3 ($S_8 = 0.759^{+0.025}_{-0.023}$ ) \cite{amon2022dark}; and $ii)$ early-time measurements, namely Planck 2018 ($S_8 = 0.834^{+0.016}_{-0.016}$) \citep{aghanim2020planck}, and different values of $S_8$ is reported from ACT collaboration, the $S_8 =0.830 \pm 0.043, 0.840 \pm 0.030$ and $0.846 \pm 0.016 $, for ACT, ACT + WMAP ACT + WMAP ACT + Planck observations respectively (more detail is presented in the work \cite{aiola2020atacama}). 
 \\
 \\For instance, the sigma difference between KiDS-1000 and Planck measurements is approximately $2.68\sigma$, indicating a tension. Our results together with the late- and early-time measurements are presented in Fig.~\ref{fig:enter-labelH0} (right panel) for all combined datasets. Using the $S_8$ values from Table~\ref{table-bestfit}, the diffusive model differs from Planck measurements by $0.56\sigma$, $-0.51\sigma$, and $-1.39\sigma$, and differs from KiDS-1000 by $0.56\sigma$, $-0.51\sigma$, and $-1.39\sigma$ for the \textit{PPS + CC + DESY5 + RSD + $f$}, \textit{DESI DR2 BAO + CC + DESY5 + RSD + $f$}, and \textit{DESI DR2 BAO + CC + Union3 + RSD + $f$} datasets, respectively. In the same way, the $\Lambda$CDM model differs from Planck measurements by $-0.33\sigma$, $-0.65\sigma$ and $-1.70\sigma$, and differs from KiDS-1000 by $1.07\sigma$, $0.94\sigma$, and $-0.06\sigma$. From this comparison we can conclude that the diffusive model's values of $S_8$ possibly form a bridge between late- and early-time measurements and has a slightly alleviating the tensions, prompting further investigation to address $S_8$ tensions. However, addressing both cosmological tensions \(H_0~\text{and}~ S_8\) more quantitatively by including the Planck 2018 data is beyond the scope of this study.

\subsection{Background evolutions}\label{backroundevolution}
The Hubble parameter, the deceleration parameter $q(z)$, the effective equation of state parameter $w_{eff}(z)$, and the distance modulus $\mu(z)$,  from Eqs \eqref{normalizedhubbel},  \eqref{deccx}, \eqref{effective}, and \eqref{distancemodulus1}, respectively, have also been presented numerically in Fig.  \ref{fig:enter-labelback} to demonstrate the background evolution of the Universe in the diffusive model. As mentioned earlier, we have considered the values of  $Q_{dm} = -0.017, -0.013,-0.009$ together with the values of  $H_0 = 68.959$, $\Omega_m = 0.361$   for the diffusive model and the corresponding values of $\Omega_m = 0.310$ and $H_0 = 69.950$ for $\Lambda$CDM model taken from Table \ref{table-bestfit}.
By emphasizing these key cosmological parameters (i.e., $q(z)\;, w_{eff}(z)\;, H(z)\;, \mu(z)$), we present the numerical results in Fig. \ref{fig:enter-labelback} to show the implications of the diffusive dark-fluid model in comparison with the $\Lambda$CDM model.  The deceleration parameter \( q(z) \), represented by the upper left of Fig. \ref{fig:enter-labelback}, shows how both models depart from a decelerating Universe to an accelerating Universe at comparable redshifts.
\\
\\
Similarly, in the upper right panel of Fig. \ref{fig:enter-labelback}, the effective equation of state \( w_{\text{eff}}(z) \) demonstrates a key difference between the two models. This dynamic behaviour could offer a more accurate explanation for the accelerated expansion observed today and potentially alter predictions about the cosmic expansion. In the bottom left of Fig. \ref{fig:enter-labelback},  the Hubble parameter \( H(z) \) of both models with the observational data is presented.  Finally, the bottom right panel of Fig. \ref{fig:enter-labelback}, which shows the distance-redshift relation \( \mu(z) \), compares predictions for the distance-redshift relation using {Pantheon+SH0ES} data. Both models are consistent with the supernova data, but minor deviations between the models at higher redshifts may hint at the diffusive dark-fluid model offering a different interpretation of the role DE played in the past.
\\
\\As a result, these plots presented in Fig. \ref{fig:enter-labelback} show that the cosmic expansion rate is sped up consistently for all cases of $Q_{dm} <0$.  Particularly, in the past cosmic evolution $z>0$, higher deviations are significantly noticeable, which indicates that the flow of energy is insightful. Conversely, when \(Q_{dm} < 0\), energy transfers from DM to DE, leading to an enhancement of cosmic acceleration. Later in section \ref{statfindern}, we shall pay attention to how the sign of $Q_{dm}$ changes the phase of the Universe.  In the current Universe ($ z=0$), the diffusive dark-fluid model has a minimal deviation from the \(\Lambda\)CDM model, and this deviation is also significantly visible at higher redshifts, exhibiting variations that suggest a distinct rate of acceleration in the early Universe. This implies that the diffusive dark-fluid model might offer new insights into how the acceleration of the Universe’s expansion evolved.

\subsection{{State finder diagnostics}}\label{statfindern}
\begin{figure}
    \includegraphics[width=1.0\linewidth]{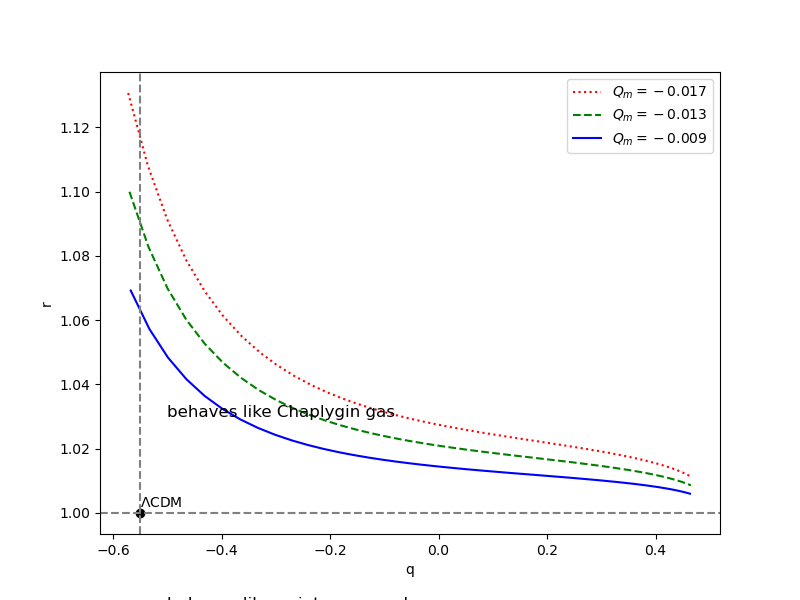}
     \includegraphics[width=1.0\linewidth]{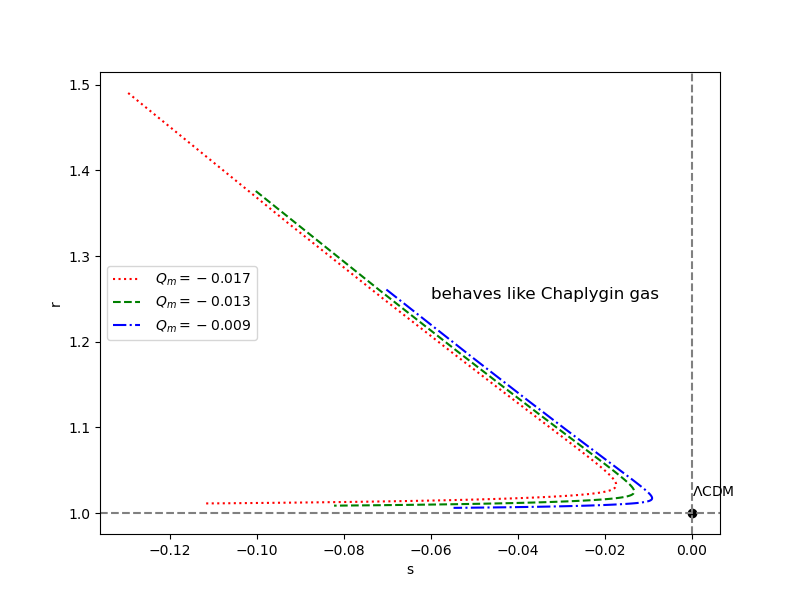}
    \caption{The state finder diagnostic \( q \) vs. \( r \) (left panel) and \( s \) vs. \( r \) (right panel).   We use $Q_{dm} = -0.017, -0.013,-0.009$, $H_0 = 68.959$ in km/s/Mpc, $\Omega_m = 0.361$   for diffusive model and $\Omega_m = 0.310$ and $H_0 = 69.950$ in km/s/Mpc for $\Lambda$CDM model.}
    \label{fig:enter-labelstatefinder}
\end{figure}

In this section, we consider the parameters for the state finder, $r$ and $s$,  which are introduced by {V. Sahni et al, 2003} \cite{sahni2003statefinder}  as  
\begin{eqnarray}
    r \equiv \frac{\dddot{a}}{aH^3} =  q(2q+1)+(1+z)\frac{dq}{dz}\;,
    \quad\mbox{and}\quad s = \frac{r-1}{3(q-\frac{1}{2})}\;,
\end{eqnarray}
respectively, for a better distinction between diffusive and $\Lambda$CDM models.  {The numerical results of these state-finder diagnostic plots are shown in Fig.  \ref{fig:enter-labelstatefinder}. The plot shows the diagram of  \( q \)  vs  \( r \) (upper panel) and \( s \) vs \( r \) (bottom panel) for different the values of \( Q_{dm} \), offer valuable insights into the nature of cosmic expansion and the role of the diffusive dark-fluid model in comparison to the standard $\Lambda$CDM model.  Based on the values of the pair \(r,s\), the diffusive dark-fluid model is categorized into the following three classes:
\begin{itemize}
    \item $r>1$, $s<0$, CG model when \(Q_{dm}<0\), 
    \item $r<1$, $s>0$, Quintessence model when  \(Q_{dm}>0\), 
    \item $r =1$, $s = 0$, $\Lambda$CDM model when \(Q_{dm} = 0\).
\end{itemize}
}

These diagnostics help to distinguish between various DE models and understand their impact on the Universe's evolution. In the \( q \) vs. \( r \) plot, the black dot refers to the $\Lambda$CDM model, representing a Universe with a cosmological constant driving its accelerated expansion.
For all cases of negative value of \( Q_{dm} \)  behaves like the CG model \cite{alam2003exploring}, which implies a late dominance of DE, causing an accelerating expansion. Here, the Universe expands at a faster rate than $\Lambda$CDM model, with \( r \)-values exceeding $1$.  Similarly, the \( s \) vs. \( r \) plot (bottom panel) shows that the CG-like model produces faster cosmic expansion, as evidenced by the negative \( s \)-values. 
\begin{figure}
    \centering
    \includegraphics[width=1.0\linewidth]{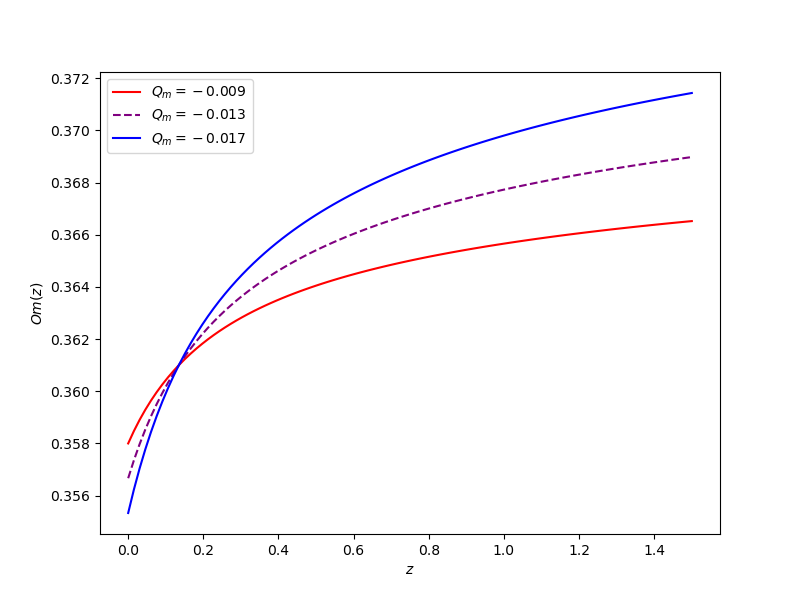}
    \caption{$Om(z)$ diagnostic diagram for the diffusive model for the case of $Q_{dm} = -0.017, -0.013,-0.009$ and $\Omega_m=0.361$.}
    \label{fig:enter-labelk}
\end{figure}
\\
\\
We also consider the $Om(z)$ diagnostics which expressed as \cite{sahni2003statefinder,sahni2008two}
 \begin{equation}
Om(z)=\frac{\left[ \frac{H(z)}{H_{0}}\right]^{2}-1 }{(1+z)^{3}-1}~. \label{eq27}
\end{equation}
 We present the numerical results of $Om(z)$ for diffusive models as presented in Fig.  \ref{fig:enter-labelk}. From our results, we have noticed that the diffusive model is similar to \(\Lambda\)CDM model, where DE behaves as a cosmological constant (where $Q_{dm} = 0$), \(Om(z)\) remains approximately constant at different redshifts. For the case of $Q_{dm} <0$, \(Om(z)\) decreases with cosmic time,  and the diffusive model behaves as CG, which leads the cosmic acceleration. 

\subsection{{Structure growth}}\label{perturbationdynamics}
{As mentioned earlier, the coupled equation of the density contrast presented in Eqs. \eqref{matter} and \eqref{cosmologicalconstant}  are taken into account  to present the numerical results of $\delta(z)$ for matter and DE using the best-fit values of $\Omega_m = 0.361$ and $Q_{dm} = -0.017, -0.013,-0.009$ taken from Table \ref{table-bestfit}  as shown in Fig. \ref{fig:enter-labeldensity}.  The value of $\delta_{de}(z)$ is generally small compared with the corresponding values of $\delta_m(z)$. The figure clearly shows \(\delta_m(z)>>\delta_{de}(z)\), highlighting the dominant role of matter perturbations in having a significant effect on gravitational collapse, which is the cause of structure formation. Hereafter, we choose the density contrast \(\delta_m(z)\) for further discussion of the growth rate $f(z)$ and the redshift space distortion $f\sigma_8(z)$.  In this Figure, the density contrast, \(\delta_m(z)\geq 1\), indicates that the density contrast grows and becomes more pronounced, leading to nonlinear behaviour. In this regime, matter clusters into dense regions, forming structures such as galaxies and galaxy clusters.} The plot shows the evolution of the matter density contrast, \(\delta_m(z)\) as a function of redshift, comparing the standard \(\Lambda\)CDM model (solid red curve) with the diffusive dark-fluid model characterized by different values of \( Q_{dm} \) (blue, green and black dashed curve). In the diffusive dark-fluid model, with parameters \(\Omega_m = 0.361\), and $Q_{dm} = -0.017, -0.013,-0.009$, there is a deviation from the \(\Lambda\)CDM model (\(\Omega_m = 0.310\)), showing significant impact enhanced structure formation. 
\begin{figure}
\includegraphics[width=1.1\linewidth]{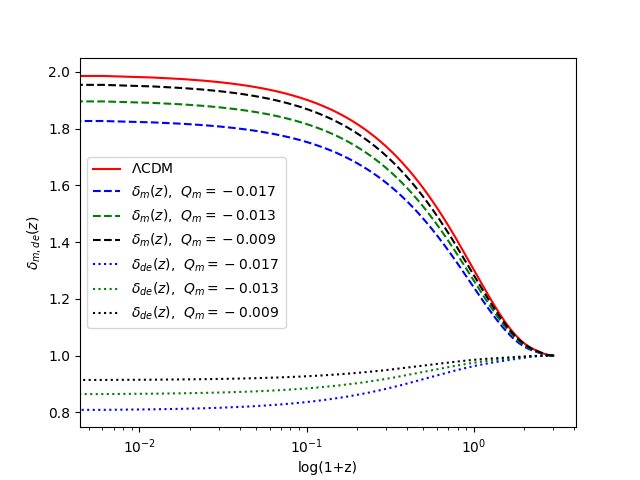}
    \caption{The numerical results of the density contrast $\delta_{{de},m}(z)$ (using the coupled system equations \eqref{matter} and \eqref{cosmologicalconstant}) for diffusive cosmology with $\Lambda$CDM models. We use $\Omega_m = 0.361$, $Q_{dm} = -0.017, -0.013,-0.009$ and $\Omega_{de} = 1-\Omega_m$.}
    \label{fig:enter-labeldensity}
\end{figure}
 Specifically, the model with a negative \(Q_{dm} = -0.017\) (blue dashed curve) shows a significant slowing in the structure formation across all redshifts, suggesting that the growth of density fluctuations is more delayed compared to the \(\Lambda\)CDM model. This demonstrates that the diffusion of matter into DE slows gravitational collapse or speeds up the accelerating cosmic expansion, resulting in a slower rate of structure growth. 
 \\
 \\
{Additionally, by accounting for the growth rate function presented in Eq. \eqref{growth11x}, we present the numerical results of $f(z)$ shown in Fig.  \ref{fig:enter-labeldensity} (upper panel ) for diffusive and $\Lambda$CDM models using $Q_{dm} = -0.017, -0.013,-0.009$. In the same manner,  the redshift space distortion data \({f}\sigma_8 \) from Eq. \eqref{growth11}
 the diagram of the $f\sigma_8(z)$ has been presented as illustrated in Fig.  \ref{fig:enter-labeldensity} ( bottom panel). These plots compare the diffusive dark-fluid model with the same values of $Q_{dm}$ along with \(\Lambda\)CDM in terms of the growth rate and the redshift space distortion \( f\sigma_8 \) diagrams. Notably, the diffusive model deviates slightly from the $\Lambda$CDM at lower redshifts and exhibits a more pronounced deviation at higher redshifts.  This suggests that the impact of energy diffusion between DM and DE has an impact on the growth of structures.
 
\begin{figure}
 \includegraphics[width=1.0\linewidth]{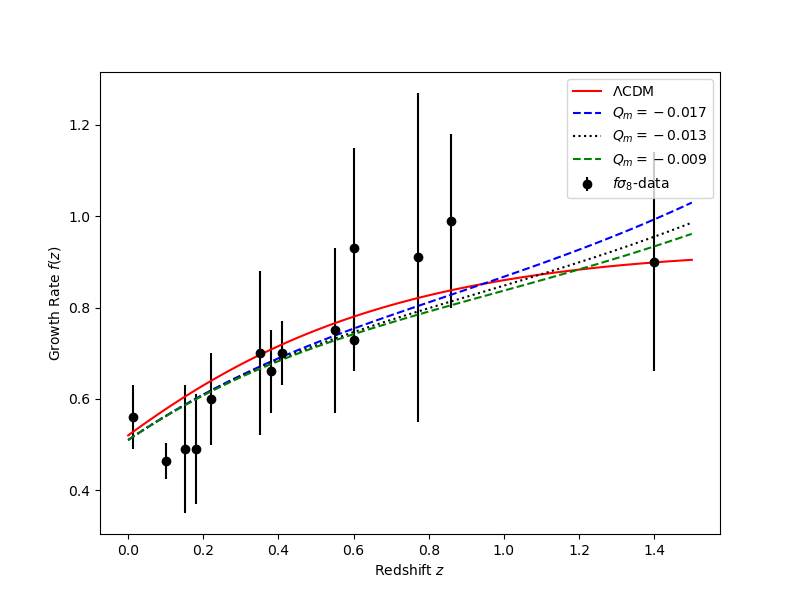}
    \includegraphics[width=1.0\linewidth]{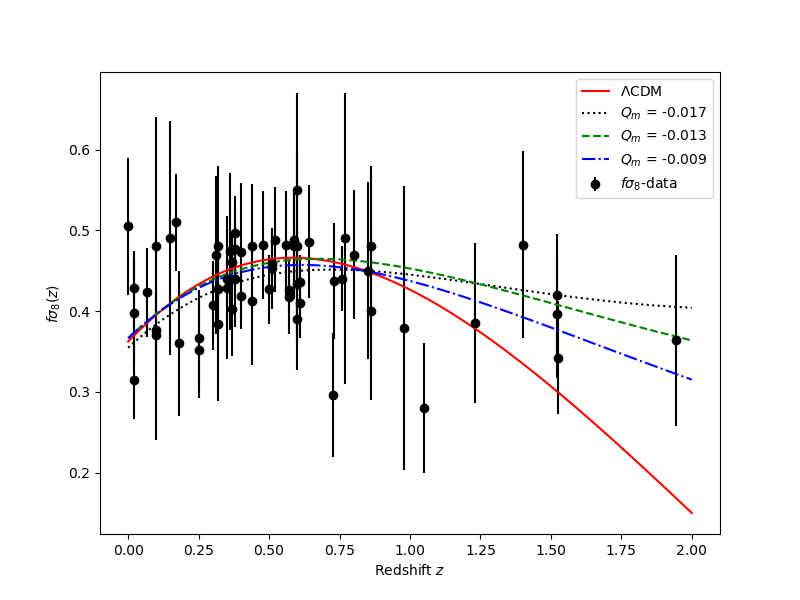}
    \caption{Upper panel:  the numerical results of the growth rate  $f(z)$ from Eq. \eqref{growth11x} for both theories: $\Lambda$CDM and diffusive dark-fluid models with observational growth rate data with error bars.  Bottom panel: the diagram of the  \( f\sigma_8(z)\)  from using Eq. \eqref{growth11}) as a function of redshift \( z \) using $Q_{dm} = -0.017, -0.013,-0.009$. The observational data points are shown with error bars.}
    \label{fig:enter-labeldensity}
\end{figure}  
 

\begin{table*}
\caption{Comparison of statistical estimators $\mathcal{L}(\hat{\Theta}|data)$, $\chi^2$, reduced $\chi^2_\nu$, AIC, $\Delta$AIC, BIC, and $\Delta$BIC between the $\Lambda$CDM and diffusive dark-fluid models for various dataset combinations.}
\label{stastical21x}
\begin{tabular}{llccccccc}
\hline
\textbf{Data Combination} & \textbf{Model} & $\mathcal{L}(\hat{\Theta}|data)$ & $\chi^2$ & $\chi^2_\nu$ & AIC & $\Delta$AIC & BIC & $\Delta$BIC \\
\hline
\multirow{2}{*}\textit{DESI DR2 BAO + CC +} 
& Diffusive & $-66.630$& 131.26& 0.924 & 143.265 & 1.322 & 158.212 & 5.690 \\
&&&&&&&\\
\textit{ Union3 + RSD + f}& $\Lambda$CDM & $-66.971 $& 133.942 & 0.936 & 141.942 & --- & 153.903 & --- \\
\hline

\multirow{2}{*}\textit{DESI DR2 BAO + CC + DESY5 } 
& Diffusive & $-919.250$ & 1838.5 & 1.049 & 1848.500& 0.410& 1875.927 & 5.120 \\
&&&&&&&\\
\textit{+ Union3 + RSD + f}& $\Lambda$CDM & $-920.456$ & 1840.912 & 1.049 & 1848.912 & ---& 1870.799 & --- \\
\hline
\multirow{2}{*}\textit{PPS + CC + DESY5 + } 
& Diffusive &$ -1698.103$ & 3396.206 & 0.987 & 3407.206& 1.060& 3436.930& 6.080\\
&&&&&&&&\\
\textit{ RSD + f} & $\Lambda$CDM &  $-1699.133$ & 3398.266 &0.987 & 3408.266 & --- & 3430.845 & --- \\
\hline
\end{tabular}

\end{table*}
\subsection{Statistical analysis}
The Bayesian/Schwarz Information Criterion (BIC) and Akaike Information Criterion (AIC) are used in our statistical analysis to evaluate the suitability of diffusive models compared to $\Lambda$CDM. As broadly described in the work \citep{liddle2009statistical,szydlowski2015aic,rezaei2021comparison}, we consider the statistical computations BIC and AIC to determine if the diffusive model should be ``accepted" or ``rejected" in comparison to $\Lambda$CDM. We use the $\Lambda$CDM as the ``accepted"  model for comparison to support our claims using the $\rm{AIC}$ and $\rm{BIC}$ criteria. By using these standards, we can determine if the diffusive model will be accepted or rejected.
The following relations are used for calculating the $\rm{AIC}$ and $\rm{BIC}$ values in the $\Lambda$CDM and diffusive models: $\rm{AIC} = \chi ^{2} +2K, ~\text{and}~ \rm{BIC} = \chi ^{2} +K\log(N_i),$ where $\chi^{2}$ is computed using the model's Gaussian likelihood function $\mathcal{L}(\hat{\Theta} |\text{data})$, the number of free parameters for that specific model is $K$. At the same time, $N_i$ is the number of data points for the $i^{th}$ dataset. Consequently, by defining the AIC Bayes factor, $$\Delta \rm{\rm{AIC}} = \big|\rm{AIC}_{\rm{\rm{Diffusive}}} - \rm{\rm{AIC}}_{\it\Lambda\rm{CDM}}\big|\;,$$ where $\rm \Delta AIC \leq 2$ indicates that the proposed theoretical model holds a \textrm{substantial observational support} for the fitted data, $ 4 \leq {\rm \Delta AIC} \leq 7$ indicates \textrm{less observational support}, and finally $\Delta \rm{AIC} \geq 10$ indicates \textrm{no observational support} as stated in \citep{szydlowski2015aic}.
The BIC Bayes factor can also be expressed as follows: $$\Delta\rm{BIC} \equiv 2\ln \rm{BIC} =  - (\rm{BIC}_{\it i} - \rm{BIC}_{j})\;,$$ where \(2\ln \rm{BIC}\) is the BIC Bayes factor comparing model (i) against model (j). In this case, (i) stands for the \(\Lambda\)CDM model, and (j) for the diffusive model. The following is a ranking of the evidence against \(\Lambda\)CDM, i.e., in favor of the diffusive model, based on the categorization in \cite{szydlowski2015aic} is negligible if \(0 \leq 2\ln \rm{BIC} \leq 2\), positive if \(2 \leq \Delta\rm{BIC} \leq 6\), strong if \(6 \leq \Delta\rm{BIC} \leq 10\), and extremely strong if \(\Delta\rm{BIC} > 10\), see Table \ref{stastical21x} for more details. Table \ref{stastical21x} displays the complete model's comparison, and the findings indicate that all values of $\Delta\rm{AIC}$ are less than 2.  According to $\Delta \text{AIC}$, this implies that the diffusive model is competitive with \(\Lambda\)CDM.  However, the values of $\Delta\rm{BIC}$ fall within the range of \(2 \leq \Delta\rm{BIC} \leq 6.080\). This suggests that there is strong evidence against \(\Lambda\)CDM in favor of the diffusive model. Based on the work in \cite{ parkinson2005testing, biesiada2007information}, due to the large number of data $N_i$ such as PPS and DESY5, AIC tends to favor models with more parameters. In contrast, BIC tends to penalize them. Even the recent work  \cite{sahlu2025structure} highlighted a similar approach, regarding the model is more penalized by BIC because of the more data points. However, the statistical preference remains moderate, meaning that while the diffusive model demonstrates potential advantages or consistency with certain observations, the level of support is not yet compelling enough to favor it decisively over  \(\Lambda\)CDM. More observational data, theoretical refinements, and robustness checks are necessary to determine whether the diffusive model can provide an explanation of cosmic dynamics compared to the  \(\Lambda\)CDM framework. 
\\
\\
The current work is an exploration of the viability of the diffusive cosmological model; it needs to be tested against more existing and upcoming data before any conclusive support for or against it is pronounced. One such data left for future work, for example, is that of the CMB, as testing the diffusive model against such data will also give us quantitative clues regarding its ability to reduce the cosmological tensions.


{ }

\section*{Acknowledgments}
AA acknowledges that this work is based on research supported in part by the National Research Foundation (NRF) of South Africa (grant number 112131). This work was part of the research programme ``New Insights into Astrophysics and Cosmology with Theoretical Models Confronting Observational Data'' of the National Institute for Theoretical and Computational Sciences of South Africa.

\section*{Data Availability}
We have used the publicly available cosmological probes, as listed in
Section \ref{resultanddiscussio}.



\bibliographystyle{mnras}
\bibliography{example}  

\end{document}